\documentclass[journal,12pt,onecolumn]{IEEEtran}
\usepackage{setspace}
\usepackage{mathpple}
\usepackage{times}

\usepackage{amsmath}  
\usepackage{amssymb}  
\usepackage{mathrsfs} 
\usepackage{arydshln}
\usepackage{theorem}  
\usepackage{cite}     
\usepackage{comment}  

\usepackage{amsfonts}
\usepackage{dsfont}
\usepackage{verbatim}
\usepackage{algorithm}
\usepackage{algorithmic}
\usepackage[dvipsnames,usenames]{color}
\usepackage{enumerate}

\usepackage{graphicx}
\usepackage{subcaption}

\usepackage{latexsym}

\usepackage{color}
\usepackage{multirow}
\usepackage{footnote}
\usepackage{booktabs}
\usepackage{cite}
\usepackage{hhline}
\usepackage{gensymb}
\usepackage{tikz}
\usepackage{graphicx}
\usepackage{xcolor}
\usepackage{listings}

\usepackage{caption}
\usepackage{mathpple}
\usepackage{times}
\usepackage{dsfont}
\usepackage{amsmath}  
\usepackage{amssymb}  
\usepackage{mathrsfs} 
\usepackage{theorem}  
\usepackage{cite}     
\usepackage{comment}  

\usepackage{amsfonts}
\usepackage{dsfont}
\usepackage{verbatim}
\usepackage{algorithm}
\usepackage{algorithmic}
\usepackage[dvipsnames,usenames]{color}
\usepackage{enumerate}

\usepackage{graphicx}

\usepackage{latexsym}

\usepackage{color}
\usepackage{multirow}
\usepackage{footnote}
\usepackage{booktabs}
\usepackage{cite}
\usepackage{hhline}
\usepackage{gensymb}
\usepackage{tikz}
\usetikzlibrary{shapes.geometric}
\usetikzlibrary{arrows.meta,arrows}
\usepackage{graphicx}
\usepackage{xcolor}
\usepackage{listings}

\usepackage{romannum}

\newcommand{\be}[1]{\begin{equation}\label{#1}}
\newcommand{\ee}{\end{equation}}

\newcommand{\bc}{\begin{center}}
\newcommand{\ec}{\end{center}}

\newcommand{\ham}[1]{d_{\{#1\}}}

\newcommand{\qed}{\hfill$\Box$\\[1ex]}

\renewcommand{\leq}{\leqslant}

\renewcommand{\geq}{\geqslant}

\newcommand{\oline}{\overline}

\newcommand{\Cref}[1]{Co\-rol\-la\-ry\,\ref{#1}}

\theoremstyle{plain} \theorembodyfont{\normalfont\slshape}

\newtheorem{thm}{Theorem$\!$}
\newenvironment{theorem}{\begin{thm}\hspace*{-1ex}{\bf.}}{\end{thm}}

\newtheorem{prop}[thm]{Proposition$\!$}

\newtheorem{lem}[thm]{Lemma$\!$}
\newenvironment{lemma}{\begin{lem}\hspace*{-1ex}{\bf.}}{\end{lem}}

\newtheorem{cor}[thm]{Corollary$\!$}

\newtheorem{prob}[thm]{Problem$\!$}

\newtheorem{defi}[thm]{Definition$\!$}

\theorembodyfont{\normalfont}

\newtheorem{exam}{Example$\!$}
\newenvironment{example}{\begin{exam}\hspace*{-1ex}{\bf .}}{\end{exam}}

\newtheorem{remrk}{Remark$\!$}
\newenvironment{remark}{\begin{remrk}\hspace*{-1ex}{\bf .}}{\end{remrk}}

\definecolor{Codecolor}{named}{White}  

\begin{document}
\title{On the Performance of Direct Shaping Codes}
\author{\large Yi Liu,~\IEEEmembership{Student Member,~IEEE,} \IEEEauthorblockN{and Paul H. Siegel,~\IEEEmembership{Life Fellow,~IEEE}}\\
	\thanks{
		Portions of this paper were presented at the IEEE Global Communications Conference, Washington, D.C., Dec. 4--8, 2016.
		
		The authors are with the Center for Memory and Recording Research, Department of Electrical and Computer Engineering, University of California San Diego, La Jolla, CA 92093, USA (e-mail: yil333@ucsd.edu; psiegel@ucsd.edu).
	}
}
\maketitle
\pagenumbering{arabic}
\thispagestyle{plain}
\pagestyle{plain}
\begin{abstract}
In this work, we study a recently proposed \emph{direct} shaping code for flash memory. This rate-1 code is designed to reduce the wear for SLC  (one bit per cell) flash by minimizing the average fraction of programmed cells when storing structured data.
Then we describe an adaptation of this algorithm that provides data shaping for MLC (two bits per cell) flash memory. It makes use of a page-dependent cost model and is designed to be compatible with the standard procedure of row-by-row, page-based, wordline programming. 
We also give experimental results demonstrating the performance of MLC data shaping codes when applied to English and Chinese language text. We then study the potential error propagation properties of direct shaping codes when used in a noisy flash device. In particular, we model the error propagation as a biased random walk in a multidimensional space. We prove an upper bound on the error propagation probability and propose an algorithm that can numerically approach a lower bound. Finally, we study the asymptotic performance of direct shaping codes. We prove that the SLC direct shaping code is suboptimal  in the sense that it can only achieve the minimum average cost for a rate-1 code under certain conditions on the source distribution.
\end{abstract}

\section{Introduction}

NAND flash memory has become a widely used data storage technology. It uses rectangular arrays, or \emph{blocks} of floating-gate transistors (commonly referred to as \emph{cells}) to store information. 
The flash memory cells gradually wear out with repeated writing and erasing, referred to as \emph{program/erase (P/E) cycling}, but the damage caused by P/E cycling is dependent on the programmed cell level. 
For example, in SLC flash memory, each cell has two different states, erased and programmed, represented by 1 and 0, respectively. Storing 1 in a cell causes less damage, or \textit{wear}, than storing 0. More generally, in multilevel flash memories, the cell wear is an increasing function of the programmed cell level. 

Many works have considered the problem of designing codes that can combat the limited-endurance in flash memory. For example, in Chee et al.~\cite{Chee}, \textit{write $l$-step-up memories} codes, which can be viewed as a generalization of non-binary \textit{write-once memory} codes~\cite{Rivest}, was proposed. This coding scheme improves the lifetime of a flash device by writing multiple messages to the flash cell before it is erased. In Liu et al.~\cite{LiuML}, a \textit{bad page detector} was introduced. This scheme improves the lifetime of a flash device by blocking the cells that wear out and store data in the remaining part of the device. 

The damage caused by programming the cell is usually modeled as a \textit{cost}, and increasing the lifetime of flash memories can be converted to the problem of encoding information for use on channels with a cost constraint. This type of code is often referred to as a \textit{shaping code}. Starting from Shannon~\cite{Shannon}, there is a substantial literature on shaping codes, for example, see Golin et al.~\cite{Golin}, Guazzo~\cite{Guazzo}, Karp~\cite{Karp} and Krause~\cite{Krause}. For pointers to more literature on shaping codes, see~\cite{LiuJournal}. Here we highlight two works that specifically address shaping codes for flash memory.
 
In~\cite{Jagmohan}, Jagmohan et al. proposed \textit{endurance coding}. For a given cost model and a specified target code rate, the optimal distribution of cell levels that minimizes the average cost was determined analytically, reproducing the results in the references cited above. For SLC flash memory, with associated level costs of 0 and 1,  greedy enumerative codes that minimize the number of cells with cost 1 were designed and evaluated in terms of the rate-cost trade-off.  However, endurance coding is intended for uniform i.i.d. source data. For structured source data, which would include a general i.i.d. source,  the idea of combining source compression with endurance coding was proposed, but the relationship between the code performance and the code rate for arbitrary sources was not thoroughly studied.

In Sharon et al.~\cite{Sharon}, low-complexity, rate-1, fixed-length \textit{direct shaping codes} for structured data were proposed for use on SLC flash memory. The  code construction used a greedy approach based upon an adaptively-built encoding dictionary that does not require knowledge of the source statistics.

In this paper, we study the properties of direct shaping codes. In Section~\ref{sec:SLC}
we review the SLC direct shaping codes proposed in~\cite{Sharon}. We illustrate its effectiveness by given an application to the  English-language novel \textit{The Count of Monte Cristo}. In Section~\ref{sec:MLC} we describe the structure and programming of MLC flash memory, and then propose a content-dependent cost model that reflects the cell wear associated with programming each level. Based on this model, we extend the data shaping code introduced in Section~\ref{sec:SLC} to MLC flash memory. We also present experimental results of applying MLC shaping codes to \textit{The Count of Monte Cristo} and Chinese-language text \textit{Collected Works of Lu Xun, Volumes 1--4}. In Section~\ref{sec:propagation}, we study the problem of error propagation by means of a random-walk recurrence analysis. We derive the upper bound on the probability of error propagation during the encoding and decoding process. We also propose an algorithm that can numerically approximate the lower bound. The analysis shows that if a sufficiently large number of codewords have been read correctly,  error propagation can be avoided. In Section~\ref{sec::performance}, we derive the asymptotic average cost of a SLC shaping code and compare the result to the asymptotically average cost of an optimal type-\Romannum{1} shaping code, which was introduced in~\cite{LiuJournal}.

\section{Shaping Codes for SLC Flash}\label{sec:SLC}

\subsection{Encoder and Decoder}

In the context of SLC flash memory, \emph{direct} shaping codes were first introduced in~\cite{Sharon}.  Their construction makes use of a rate-1,  adaptive encoding dictionary $\mathbf{D}$ that  is used to map successive words of length $m$ in the input sequence  $\mathbf{W}$ into codewords of length $m$. Denote by $\mathbf{w_i}\in \{0,1\}^m$, $i = 1,\ldots 2^m$, the input word and by $P_i$ its probability. Without loss of generality, we assume that $P_1 \geq P_2 \geq \ldots \geq P_{2^m}$. The dictionary $\mathbf{D}$  comprises two lists, a dynamically ordered list  $\mathcal{X}$ and a fixed output codeword list $\mathcal{Y}$.   The output list $\mathcal{Y}$ consists of codewords $\mathbf{y_k}\in \{0,1\}^m$,  $k=1,2,\ldots,2^m$,  ordered by non-decreasing \emph{cost}, where the cost of a codeword $\mathbf{y_k}$ in this context means the number of 0 symbols in $\mathbf{y_k}$.

At any given time during the encoding process, the list  $\mathcal{X}$ consists of pairs  $\{(\mathbf{x}_k, n_k)\}$, $k=1, 2 \ldots, 2^m$, where  each  pair  represents a distinct length-$m$ input word $\mathbf{x}_k\in \{0,1\}^m$ and the \emph{frequency count}, or number of times, $n_k$, that it has appeared up to that point in the input data sequence. The list of pairs is dynamically ordered such that the $n_k$ values are in non-increasing order, i.e, $n_1 \geq n_2  \geq ... \geq n_{2^m}$.  At the start of the encoding process, the  $n_k$  are all initialized to the value 0, and the words $\mathbf{x}_k$ are ordered lexicographically. 

Encoding proceeds as follows. When a data word  $\mathbf{x}$ of length $m$ is encountered in the input data sequence, its corresponding pair $(\mathbf{x}_k, n_k)=(\mathbf{x}, n)$ is found in the ordered list $\mathcal{X}$.  The encoder then maps 
$\mathbf{x}$ to the  length-$m$ output word $\mathbf{y}$ that occupies the same position in output list $\mathcal{Y}$.  The frequency count $n$ of the word $\mathbf{x}$ is increased by 1 and the list $\mathcal{X}$ is reordered accordingly, with the pair 
$(\mathbf{x}, n+1)$ moved above all pairs with counts less than or equal to $n+1$.

\begin{example}
	Consider a direct shaping code with parsing length $m=2$.  In the encoding dictionary $\mathbf{D}$, the ordered output list $\mathcal{Y}$ is $\{11,10,01,00\}$. Consider the length-14 data sequence $\mathbf{w} =10.11.00.10.11.10.00$.  The first six length-2 input words   contain three words $\mathbf{x}_1=10$, two  words $\mathbf{x}_2=11$ and one word $\mathbf{x}_3=00$. The state of the dictionary $\mathbf{D}$ after encoding these 6 words is shown in Table \ref{table::beforeencoding}. 

The final input word is $\mathbf{x}=00$. According to Table \ref{table::beforeencoding}, it is mapped to output codeword $\mathbf{y_3}=01$. We then add 1 to the count $n_3$ in the table and update the ordering of the entries in the input list $\mathcal{X}$. The updated  dictionary $\mathbf{D}$ is shown in Table~\ref{table::afterencoding}.\qed
\end{example}

\begin{figure}[h]
\centering
\includegraphics[width=0.6\columnwidth]{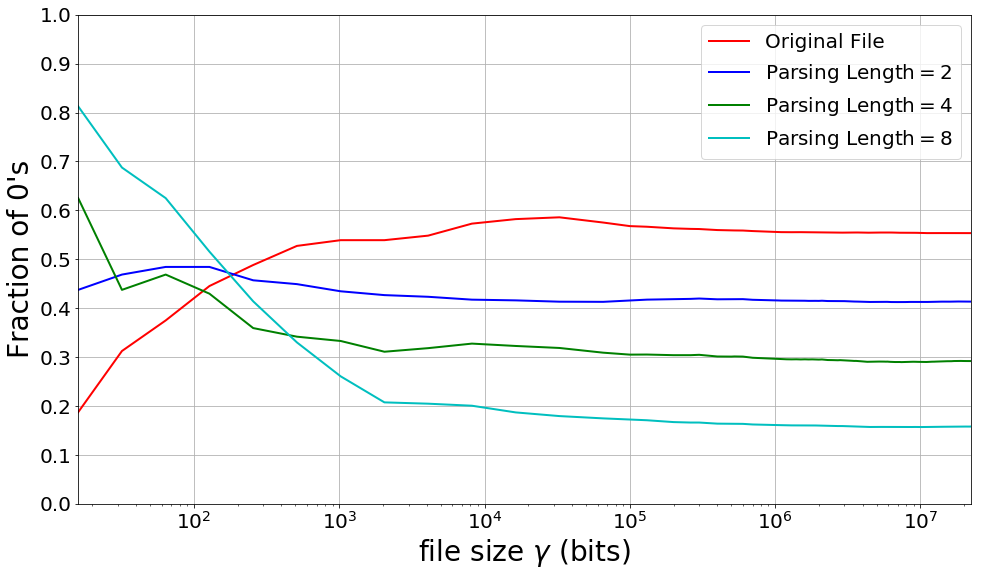}
\caption{Direct shaping codes applied to \textit{The Count of Monte Cristo}.}
\label{fig::dscTCMC}
\end{figure}

\begin{table}[h]
\centering
\begin{subtable}{0.45\columnwidth}
	\centering
	\caption{}
	\label{table::beforeencoding}
       \begin{tabular}{@{}ccccc@{}}
		\toprule
		$\mathbf{x}$ & &n  && $\mathbf{y}$ \\ \midrule
		10   &   & 3   &  & 11                     \\
		\textcolor{blue}{11}    && 2 && 10                              \\
		\textcolor{red}{00}    & &\textcolor{red}{1}   && \textcolor{red}{01}                       \\
		01     && 0      && 00                 \\\bottomrule
	\end{tabular}      
\end{subtable}~
\begin{subtable}{0.45\columnwidth}     
\centering
\caption{}
	 \label{table::afterencoding}
       \begin{tabular}{@{}ccccc@{}}
		\toprule
		$\mathbf{x}$ & &n   && $\mathbf{y}$ \\ \midrule
		10    &   & 3      &  & 11                 \\
		\textcolor{red}{00}   && \textcolor{red}{2}      && \textcolor{red}{10}                          \\
		\textcolor{blue}{11}     & &2          && 01               \\
		01  && 0    && 00                 \\\bottomrule
	\end{tabular}   
\end{subtable}
\caption{Length-2 dictionary when encoding $\mathbf{x}=00$.}
\end{table}
The decoder  dynamically  reconstructs the table and inverts the encoder mapping. When a codeword $\mathbf{y}$ is encountered in the encoded sequence, it is mapped to the binary length-$m$ data word $\mathbf{x}$ that occupies the same position in the  input list $\mathcal{X}$. Then the frequency count $n$ of the word $\mathbf{x}$ is increased by 1 and the ordering of the input list $\mathcal{X}$  is updated accordingly.  This rate-$1$ shaping code incurs no rate penalty and both encoding and decoding can be implemented with low complexity.

\subsection{Simulation Results}

We now present simulation results quantifying the endurance gain which can be achieved by the use of SLC direct shaping codes. The structured data we used was the English-language  novel \textit{The Count of Monte Cristo} (TCMC), represented in ASCII Code. We evaluated the shaping code using encoder parsing length $m$ equal to 2, 4 and 8. Fig.~\ref{fig::dscTCMC} shows the fraction of 0 symbols in the first $\gamma$ bits in the original data file and in the corresponding encoded files. The fraction of 0's in the entire original data file was approximately 0.55. With parsing length equals to 2 bits, the fraction of 0's dropped to about 0.41. With parsing length equal to 4 bits, the fraction of 0's dropped to about 0.29, and with parsing length of 8 bits, the fraction was further reduced to about 0.16.

\section{Data Shaping Codes for MLC Flash}\label{sec:MLC}

\subsection{Cost Model for MLC Flash}

\begin{figure}[h]
\centering
\includegraphics[width=0.6\columnwidth]{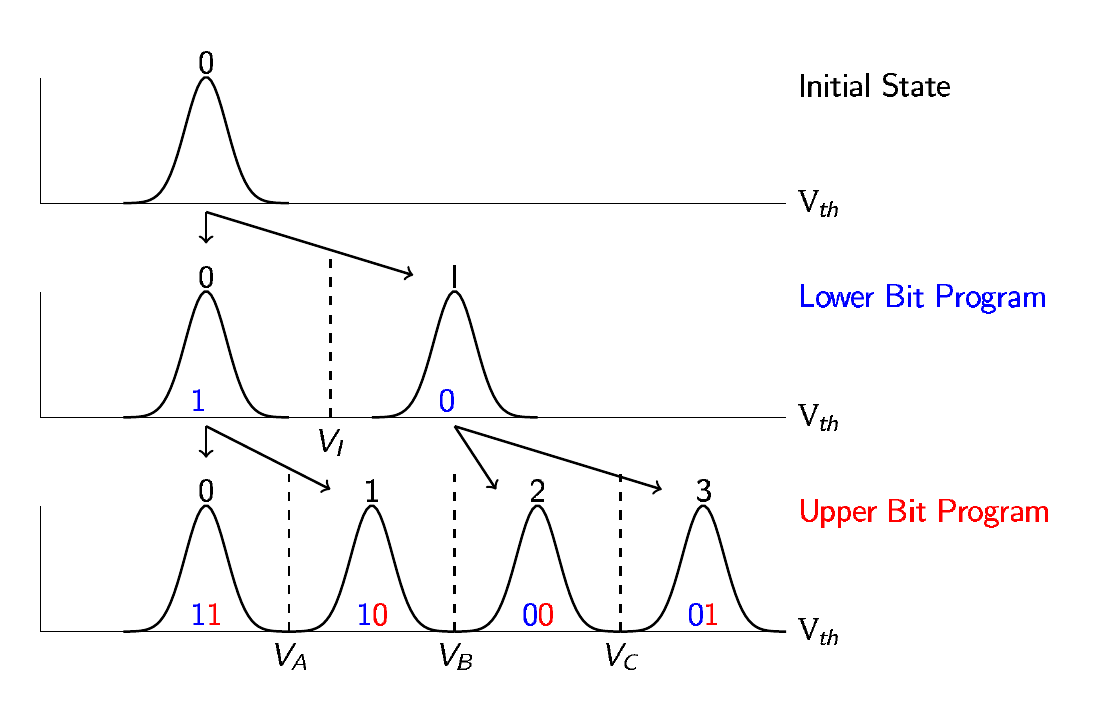}

\caption{Schematic of MLC flash cell programming.}
\label{fig:mlc_programming}
\end{figure}

Every cell in an MLC flash memory can be charged to 4 different values of the threshold voltage, $V_{th}$. Thus, each cell can represent 2 bits of information. The levels  are denoted by $0,1,2,3,$ respectively, from lowest to highest, and the corresponding binary representations are given by the Gray code $\{11, 10, 00, 01\}$. In the binary representation of a level, the left-most bit is called the \emph{lower bit} and the right-most bit is called the \emph{upper bit}. To program the MLC flash cell, we assume the controller uses a two-step process. It first charges the cell to an intermediate voltage level that reflects the value of the lower bit. Then, taking into account the value of the upper bit, it completes  charging of  the cell to reach the appropriate final threshold voltage level. This process is  shown schematically in Fig. \ref{fig:mlc_programming}.

The rows of cells in a flash memory block are called \emph{wordlines}, and the wordlines are programmed sequentially, in a row-by-row manner.   The lower bits of cells in a wordline constitute the \emph{lower page}, while the upper bits form the \emph{upper page}.   
When programming a block, information is programmed  separately to lower pages and upper pages.  During data retrieval, pages are also retrieved independently,  with lower bit values recovered by reference to read threshold $V_B$, and upper bit values by reference to read thresholds $V_A$ and $V_C$, also shown in Fig. \ref{fig:mlc_programming}.

\begin{figure}
	\centering
	\includegraphics[width=0.6\columnwidth]{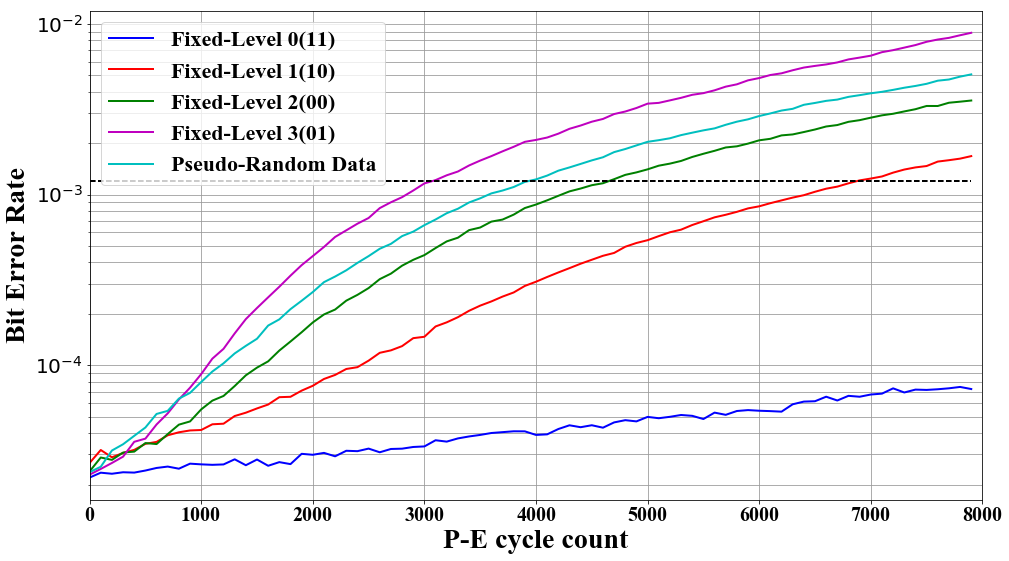}
	\caption{Cell error rate of MLC Flash for different programming levels, while repeatedly programming a constant level.}
	\label{fig::mlccer}

\end{figure}

To characterize and quantify the damage caused by different programmed levels in an MLC flash memory, we performed an experiment on several blocks in which we repeatedly programmed all the cells in the block to a specified, constant level, erasing the block (i.e., reducing voltage to level 0) after each programming cycle. The device we used was a 1x nm MLC flash chip with the default read threshold positions. After every $100$ of these P/E cycles, we programmed the block with pseudo-random data, read it back, and recorded the cell error rate. The error rates, averaged over the blocks used in the experiment, are shown in Fig. \ref{fig::mlccer}. We see that cell damage caused by the levels   $0,1,2,3$  increases monotonically.  A  \emph{cost model}, in the form of a vector of  cell-level costs $\mathcal{C}=[c_0, c_1, c_2, c_3]$, is used to quantify the relative amount of device wear associated with each of the cell levels.  In practice, these costs will satisfy $c_0 \leq c_1 \leq c_2 \leq c_3$, reflecting the increased damage induced by higher programmed levels.

Given a length-$m$ cell-level codeword $\mathbf{z}=[z_1, \ldots, z_m]$, we denote the cost associated with the symbol $z_i$ by $c(z_i)$, and
 the total cost $c(\mathbf{z})$ associated with programming $\mathbf{z}$ is assumed to be the sum of the individual symbol costs;
 i.e., $C = c(\mathbf{z})=\sum_{i=1}^m c(z_i)$.

\begin{remark}
	\label{rmk::costmodel}
	A method for determining physically meaningful values of $c_i$  was proposed by Li et al. in \cite{Li}.  It is based upon measuring the number of cycles it takes for the error rate associated with a certain programmed level to reach a prespecified \emph{maximum tolerable cell error rate}. If the design lifetime of the flash memory is $T_{0}$ P/E cycles, the maximum tolerable cell error rate $CER_{max}$ is defined as the cell error rate after the memory is programmed with random data for $T_0$ P/E cycles.  We view the damage caused by programming random data to the memory as being proportional to $1/T_0$.
	
	We define $\Phi^{i}(T)$ to be the cell error rate observed after $T$ cycles of programming the blocks to the level $i$, for $i = 0,1,2,3$.
	Let $T^{i}_{max}$ denote the P/E cycle number at which $\Phi^{i}(T)$ equals $CER_{max}$.  
	We view  the damage caused by programming to level $i$ as proportional to $1/T^i_{max}$, so we define the cost associated with each   level $i$, $i=0, 1,2, 3$, to be 
	\begin{equation}
	c_i=\frac{T_0}{T^i_{max}}.
	\end{equation}
	We calculate the cost model corresponding to the error rate results in Fig.~\ref{fig::mlccer}, assuming a design lifetime of $T_{0}=4000$ cycles. The error rate associated with level 0 remains essentially constant, so we set $c_0=0$. For the other levels, we find that  $T^{1}_{max}=6900$, $T^{2}_{max}=4600$ and $T^{3}_{max}=3100$, so the complete cost model is computed to be $[0,0.58,0.87,1.29]$.\qed
\end{remark}

\subsection{Shaping Codes for MLC Flash}

Application of the SLC shaping code independently to lower and upper pages will not be effective in reducing the average cost. This can be seen by  referring to the schematic of the MLC flash cell programming process in Fig. \ref{fig:mlc_programming}.  As shown in the schematic, the cost associated with an upper bit 1 or 0 depends on the value of the  corresponding lower bit in the cell.  The  proposed MLC shaping encoder achieves improved wear reduction by  using lower-page dependent dictionaries when encoding the upper pages, as we now describe.

First, fix a parsing length $m$.  Encoding and programming of wordlines is done in a row-by-row, sequential manner. Suppose that we want to encode $L$ data words to both lower and upper pages. For the lower pages, the encoder simply uses the direct shaping code with parsing  length $n$.  When programming an upper page, however, the encoder first reads the corresponding, previously programmed lower page. 

Now, suppose the lower page that has been programmed and presumably correctly recovered  consists of the sequence of length-$m$ codewords $\bf{v}^{(k)}$, $k=0, \ldots, L$.  Consider the sequence of length-$m$ data words $\bf{w}^{(k)}$, $k=0, \ldots, L$   that need to be encoded for the upper page.  To encode the $k$th data word $\bf{w}^{(k)}$, we encode using a direct shaping code that is  based upon an adaptively-built dictionary that depends on the corresponding lower page codeword $\bf{v}^{(k)}$.  The only difference in the operation of each lower-page dependent shaping encoder is that the ordering of the length-$m$ encoder output words in terms of increasing cost depends specifically on the lower page codeword $\bf{v}$ and the cost model $[c_0, c_1, c_2, c_3]$. 

To illustrate the design of the encoder, we will use the cost model  $[c_0, c_1, c_2, c_3]=[0, 1, 1, 2]$. 
This simple cost model is motivated as follows. Consider the standard lower-upper page binary representation of the cell levels: 
0=11, 1=10, 2=00, 3=01. We assign a cost of 0 to lower bit value 1, and a cost of 1 to  lower bit value  0, in accordance with the MLC cell programming schematic.  When the lower bit value is 1, we assign a cost of 0 to upper bit value 1, and a cost of 1 to upper bit value 0.  On the other hand, when the lower bit value is 0, we assign a cost of 0 to upper bit value  0, and a cost of 1 to upper bit value  1.  Again, these cost assignments are consistent with the MLC cell programming schematic. The cost associated with a cell level is then defined as the sum of the corresponding lower bit and upper bit costs.

Choose parsing  length $n=4$ and assume the lower page codeword is $\bf{v}=1110$.  For the first three bits, where the corresponding lower bit is a 1, programming the upper bit to $1$ is better than to $0$.  For the last bit, where the  corresponding lower bit is a $0$, programming  the upper bit to $0$ is better than to $1$.  This implies that the lowest cost output word in the dictionary is $1110$, corresponding to cell level word $0002$, which has total cost 1.  Similar reasoning leads to the ordered list of output words shown in Table \ref{table::upperdictionary_1110}. 
The corresponding list of cell level words is given in Table \ref{table::upperdictionary_1110_cells}. 
\begin{table}
	\centering
	\begin{tabular}{@{}cccc@{}}
		\toprule
		index & Output List & index & Output List \\ \midrule
		0     & 1110              & 8      & 1000              \\
		1     & 1111              & 9      & 0100                \\
		2     & 1100              & 10    & 0010               \\
		3     & 1010              & 11    & 1001            \\ 
 		4     & 0110              & 12    & 0101  \\
 		5     & 1101	      & 13    & 0011\\
		6     & 1011              & 14    & 0000 \\
		7     & 0111              & 15    & 0001   \\\bottomrule
	\end{tabular}
\caption{Ordered list of upper page words for  lower page codeword $1110$.}
\label{table::upperdictionary_1110}
\end{table}

\begin{table}
	\centering
	\begin{tabular}{@{}cccc@{}}
		\toprule
		index & Output List & index & Output List \\ \midrule
		0     & 0002            & 8     & 0112             \\
		1     & 0003             & 9     & 1012               \\
		2     & 0012            & 10    & 1102             \\
		3     & 0102            & 11    & 0113               \\ 
		4     & 1002            & 12    & 1013    \\
		5     & 0013            & 13    & 1103  \\
		6     & 0103            & 14    & 1112  \\
 		7     & 1003            & 15    & 1113  \\\bottomrule
	\end{tabular}
	\caption{ Corresponding list of cell level words for lower page codeword $1110$.}
\label{table::upperdictionary_1110_cells}
\end{table}
It is easy to verify that the  costs of the corresponding cell level words are non-decreasing:  cost 2 for words 1 to 4, cost 3 for words 5 to 10, cost 4 for words 11 to 14, and cost 5 for word 15.  In general, the encoder uses $2^m$ direct shaping encoders operating in a sequence determined by the sequence of lower page codewords $\bf{v}^{(k)}$, $k=0, \ldots, L$.

\subsection{Encoding Algorithm for MLC Shaping Codes}
The upper page encoder uses an encoding table in which the order of the output codewords depends on the lower page codeword. The encoding table provides a mapping from an index $I \in \{1, \ldots, 2^m\}$ to the corresponding upper page output word $\mathbf{y}$ in the ordered list. This mapping can be implemented using enumerative coding, introduced by Cover \cite{Cover}. To explain the algorithm, we first introduce some notation. We denote  the length-$m$ lower page codeword by $\mathbf{v}$ and  the  upper page word by $\mathbf{y}$.  They determine the cell level output word $\mathbf{z} = [z_1, \ldots, z_m]$, where $z_i$ is the cell level associated with the lower and upper bit pair $v_i y_i$. 

Let $n(\mathbf{v},C,y_1,y_2,\ldots,y_k)$ be the number of upper page words $\mathbf{y}$ that, together with lower page codeword  $\mathbf{v}$, yield  total cost $C$, and whose first k coordinates are equal to $[y_1,y_2,\ldots,y_k]$.  To determine $n(\mathbf{v},C,y_1,y_2,\cdots,y_k)$, we first calculate the polynomial $g_{\mathbf{v},k}(x)=(x^{c_0}+x^{c_1})^{n_1}(x^{c_2}+x^{c_3})^{n_0}$, where $n_b$, $b\in\{0,1\}$ is the number of bits equal to $b$ in the last $m-k$ bits of the lower page codeword, $[v_{k+1}, \ldots, v_m]$.  The degrees of terms in $g_{\mathbf{v},k}(x)$ represent the possible total costs of cell level  vectors $[z'_{k+1}, \ldots, z'_m]$  with associated lower bit vector $[v_{k+1}, \ldots, v_m]$, and the corresponding coefficients represent the number of such vectors. 
Then $n(\mathbf{v},C,y_1,y_2,\cdots,y_k)$ is equal to the coefficient of the term in $g_{\mathbf{v},k}(x)$ whose degree is equal to $C-\sum_{i=1}^{k}u_i$.  If there is no such term in the polynomial, we set $n(\mathbf{v},C,y_1,y_2,\cdots,y_k)=0$. In particular, the  polynomial $g_{\mathbf{v},0}(x)$ corresponding to $[v_1, \ldots, v_m]$ tells us the costs  $C_j, j=1, \ldots, J$ of all possible length-$m$ cell level vectors with associated lower bit vector $\mathbf{v}$, along with the number of  upper page words  that together with $\mathbf{v}$ produce these costs. From these we calculate $n(\mathbf{v},C)$, the number of upper page words that produce total cost less than or equal to U, given lower page codeword $\mathbf{v}$.   Example \ref{example::calculaten} illustrates how to calculate  $n(\mathbf{v},C,y_1,y_2,\cdots,y_k)$. 

\begin{example}\label{example::calculaten}
	We want to calculate $n(\mathbf{v},C,1,1)$ where the lower page codeword is $\bf{v}=1110$ and the total cost is $C=2$. The combined cost of the first two cells is 0, so the remaining cost is 2. We calculate the polynomial $(x^{c_0}+x^{c_1})^{n_1}(x^{c_2}+x^{c_3})^{n_0}= (1+x)(x+x^2)=x+2x^2+x^3$. The coefficient of $x^2$ is 2, meaning there are two possible codewords with total cost $C=2$ and first two bits  $[1,1]$.\qed\end{example}

With enumerative coding,  instead of storing the ordered upper page output word list, we only need to store the values $n(\mathbf{v},C,y_1,y_2,\cdots,y_k)$ and $n(\mathbf{v},C)$.  Algorithm~\ref{algo::directencoding} describes the encoding process.
\begin{algorithm}[ht]
	\caption{Encoding upper page codewords}
	\label{algo::directencoding}
	\algsetup{
		linenosize=\small
	}
	\begin{algorithmic}[1]
		\REQUIRE Codeword length $m$, lower page codeword $\bf{v}$, index $I$
		\ENSURE Upper page codeword $\mathbf{y}=(y_1,y_2,\ldots,y_m)$
		\STATE Find $j$ such that $I>n(\mathbf{v},C_{j-1})$ and $I\leq n(\mathbf{v},C_{j})$, set $I = I - n(\mathbf{v},C_{j-1})$
		\STATE If $I>n(\mathbf{v},C_j,0)$ set $y_1=1$ and set $I=I-n(\mathbf{v},C_j,0)$, otherwise set $y_1=0$
		\STATE For $i$th bit, if $I>n(\mathbf{v},C_j,y_1,\cdots,y_{i-1},0)$ set $y_i=1$ and set $I=I-n(\mathbf{v},C_j,y_1,\cdots,y_{i-1},0)$, otherwise set $y_i=0$
	\end{algorithmic}
\end{algorithm}

\subsection{Simulation Results}
\label{subsec::seperateMLC_sim}
We assessed the performance of the MLC shaping encoder using TCMC.
The data file was divided into two consecutive subfiles of size $11,134,796$ bits; the first was used for lower page encoding and the second for upper page encoding.  

\begin{figure}[h]
	\centering
	\includegraphics[width=0.6\columnwidth]{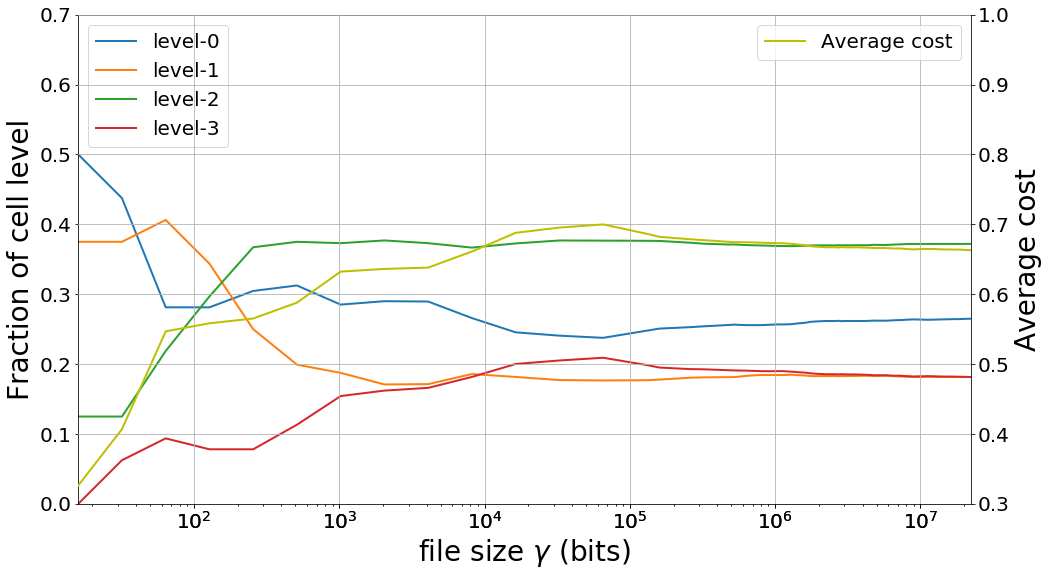}
	\caption{Fractions of MLC cell levels for segments of \emph{The Count of Monte Cristo} without a shaping code.}
	\label{fig::TCMC_mlc_no_shaping}
\end{figure}

\begin{figure}[h]
	\centering
	\includegraphics[width=0.6\columnwidth]{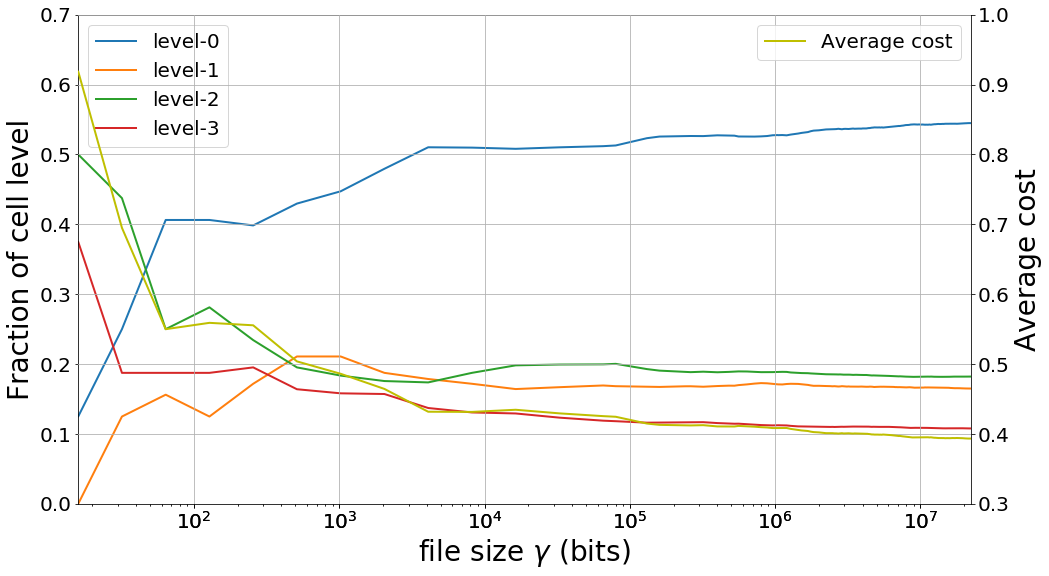}
	\caption{Fractions of MLC cell levels for segments of \emph{The Count of Monte Cristo} with MLC shaping code.}
	\label{fig::TCMC_mlc_shaping}
\end{figure}

Fig.~\ref{fig::TCMC_mlc_no_shaping} shows the respective fractions of the encoded cell levels $0,1,2,3$ appearing in the cell level sequences produced when no shaping code was applied to the first $\gamma$ bits of each subfile. Fig.~\ref{fig::TCMC_mlc_shaping}  shows the corresponding fractions when the proposed MLC shaping encoder is applied using the empirical cost model derived in Remark.~\ref{rmk::costmodel}.


From these figures, we see that for TCMC, the average cost when no coding is applied is 0.66.  When SLC shaping is applied independently to lower and upper pages, the average cost is 0.48 (not shown). In contrast, after MLC shaping, the average cost is reduced to  0.39.  

\subsection{Experimental Results on MLC Flash Memory}
\begin{figure}[h]
	\centering
	\includegraphics[width=0.6\columnwidth]{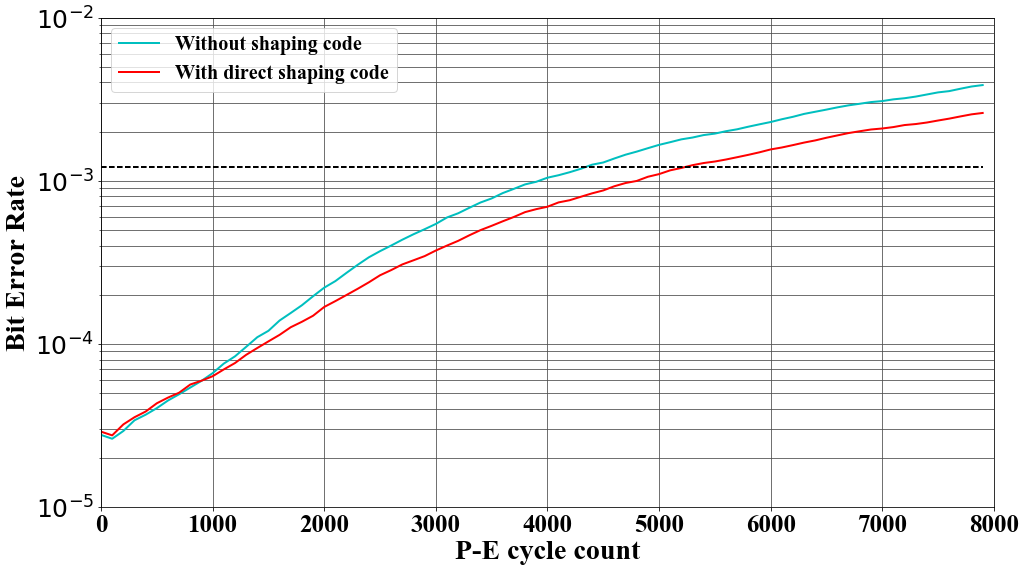}
	\caption{BER performance of English-language text \textit{The Count of Monte Cristo}.}
	\label{fig::TCMC}
\end{figure}

\begin{figure}[h]
	\centering
	\includegraphics[width=0.6\columnwidth]{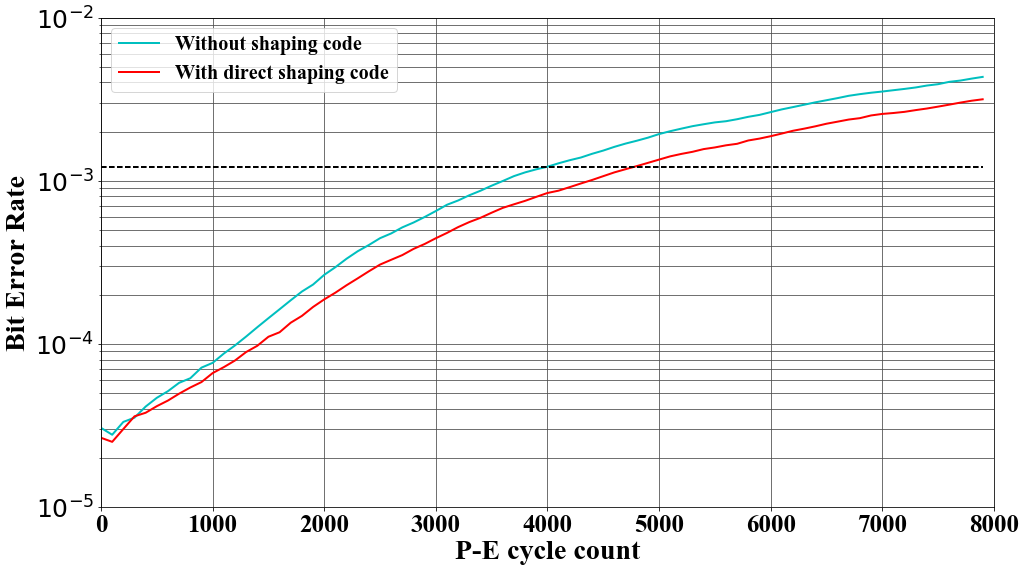}
	\caption{BER performance of Chinese-language text \textit{Collected Works of Lu Xun, Volumes 1--4}.}
	\label{fig::LUXUN}
\end{figure}

In this subsection, we present the experimental results that evaluate the performance of MLC shaping codes.
To characterize the performance of the shaping code, we performed a program/erase (P/E) cycling experiment on the MLC flash memory by repeating the following steps, which collectively represent one P/E cycle.
\begin{itemize}
	\item Erase the MLC flash memory block.
	\item Program the MLC flash memory.
	\item For each successive programming cycle, ``rotate" the data, so the data that was written on the $i$th wordline is  written on the $i+1$st wordline, wrapping around the last wordline to the first wordline.
	\item After every 100 P/E cycles, erase the block and program pseudo-random data. Then perform a read operation, record bit errors, and calculate the bit error rate.
\end{itemize}
The experiment was conducted with the uncoded source data, and then with the output data from the MLC shaping code described in Section~\ref{subsec::seperateMLC_sim}. The average bit error rates (BERs) are shown in Fig.~\ref{fig::TCMC}. The black dash line represents the maximum tolerable cell error rate introduced in Remark~\ref{rmk::costmodel} and the increased lifetime was measured at this rate.
The results indicate that applying MLC shaping codes to English text increases the lifetime of flash memory device by 900 P/E cycles. A similar experiment was done using the Chinese text \textit{Collected Works of Lu Xun, Volumes 1--4}. Fig.~\ref{fig::LUXUN} shows that MLC shaping codes increase the lifetime of flash memory device by 800 P/E cycles.

\section{Error Propagation Analysis}
\label{sec:propagation}
The direct shaping decoder reproduces the dynamic construction of the encoding list $\mathcal{X}$. Errors in reading the flash memory can lead to incorrect word frequency counts that, in turn, can cause decoding errors if word counts are not sufficiently separated.
In this section we introduce a framework for analyzing potential error propagation properties of direct shaping codes, based upon recurrence properties of random walks.

\subsection{Recurrence probability of a two-word dictionary}
\begin{figure}[t]
	\centering
	\begin{subfigure}[b]{0.3\columnwidth}
		\includegraphics[width=\linewidth]{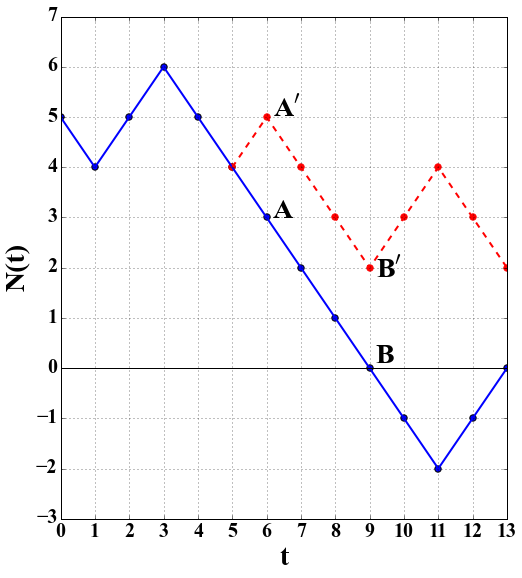}
		\caption{} \label{fig::errormodel1}
	\end{subfigure}
	\begin{subfigure}[b]{0.3\columnwidth}
		\includegraphics[width=\linewidth]{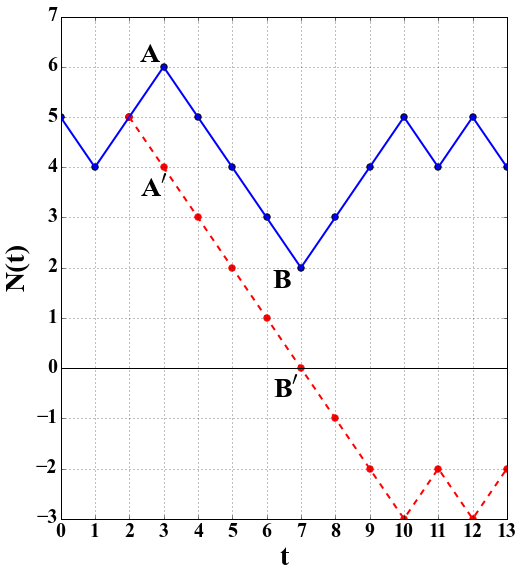}
		\caption{} \label{fig::errormodel2}
	\end{subfigure}
	\caption{Error propagation examples.}
	\label{fig::errormodel}
\end{figure}

We first consider the case when $m=1$, where the input dictionary contains only two words $\mathbf{w}_1$  and $\mathbf{w}_2$.
Suppose that, at time $t_0$, the  input word counts satisfy $n_1(t_0)\ne n_2(t_0)$. Since there are only 2 words, the dictionary after time $t$ can be represented by the word that has the higher count, denoted by $s(t)\in \{\mathbf{w}_1,\mathbf{w}_2\}$, and the distance, denoted simply by $N(t)=n_1(t)-n_2(t)$.  Let $s^e(t), N^e(t)$  represent the dictionary evolution during the encoding process, and let  $s^d(t), N^d(t)$ represent the dictionary evolution during decoding process. The distance behaves like a one-dimensional random walk, as depicted in Fig.~\ref{fig::errormodel1}. In the figure, the blue line represents the encoding process, starting at time denoted as $t=0$, with $s^e(0)=\mathbf{w}_1$ and $N^e(0)=5$.

Assuming no read errors through time $t=5$, the decoder correctly reconstructs the encoding dictionary, retracing the blue line. At time $t=5$, we have $s^d(5)=s^e(5)=\mathbf{w}_1$ and  $N^e(5)=N^d(5)=4$. At $t=6$, the input was $\mathbf{w}_2$, implying $s^e(6)=\mathbf{w}_1$ and $N^e(6)=3$,  producing output codeword 0. This point is marked by $A$ on the blue line. Now, suppose a read error occurs at decoding time $t=6$.  The codeword is incorrectly read as 1, which is decoded as input $\mathbf{w}_1$, resulting in $s^d(6)=\mathbf{w}_1$ and  $N^d(6)=5$. This deviation is marked by $A'$ on the red dashed line, which represents the decoding process. Assume no further read errors occur.  The decoder trace remains separated from the encoder trace, but the following three codewords are nevertheless decoded correctly. At time $t=9$, since $N^e(9)=0$, the encoder changes to $s^e(9)=\mathbf{w}_2$, while $N^d(9)=2$ implies that the decoder continues with $s^d(9)=\mathbf{w}_1$. These two points are indicated by $B$ and $B'$ on the encoding and decoding traces, respectively.  The next input word was $\mathbf{w}_2$  and the encoder output was $1$, but, even though it is read correctly, the decoded  word is $\mathbf{w}_1$. In fact, after this point, all codewords will be decoded incorrectly. This example shows that if  the encoding dictionary reaches a point where $N^e(t)=0$, there is the potential for error propagation. Fig. \ref{fig::errormodel2} also shows the potential of error propagation if the decoding dictionary reaches a point where $N^d(t)=0$. The encoding and decoding processes in Fig.~\ref{fig::errormodel} are summarized in Tables~\ref{tab::errormodel1} and~\ref{tab::errormodel2}.

\begin{table}
	\centering
	\scalebox{0.7}{
	\begin{tabular}{ccccccccccccccc}
		\toprule
		$t$&  0 & 1 & 2 & 3 & 4 & 5 & 6 & 7 & 8 & 9 & 10 & 11 & 12 & 13    \\\midrule
		$(s^e, N^e)$ & $(\mathbf{w}_1, 5)$ & $(\mathbf{w}_1, 4)$ & $(\mathbf{w}_1,5)$ &  $(\mathbf{w}_1,6)$  & $(\mathbf{w}_1,5)$  & $(\mathbf{w}_1,4)$ & $(\mathbf{w}_1,3)$ & $(\mathbf{w}_1,2)$ & $(\mathbf{w}_1,1)$ & $(\mathbf{w}_2,0)$ & $(\mathbf{w}_2, -1)$ & $(\mathbf{w}_2,-2)$ & $(\mathbf{w}_2,-1)$ & $(\mathbf{w}_1,0)$\\
		input & --- & $\mathbf{w}_2$ & $\mathbf{w}_1$ & $\mathbf{w}_1$ & $\mathbf{w}_2$ & $\mathbf{w}_2$ & $\mathbf{w}_2$ & $\mathbf{w}_2$& $\mathbf{w}_2$ & $\mathbf{w}_2$ &$\mathbf{w}_2$ &$\mathbf{w}_2$ &$\mathbf{w}_1$ &$\mathbf{w}_1$ \\
		output &  --- & $0$ & $1$ & $1$ & $0$ & $0$ & $0$ & $0$& $0$ & $0$ &$1$ &$1$ &$0$ &$0$\\\midrule
		$(s^d, N^d)$ & $(\mathbf{w}_1, 5)$ & $(\mathbf{w}_1, 4)$ & $(\mathbf{w}_1,5)$ &  $(\mathbf{w}_1,6)$  & $(\mathbf{w}_1,5)$  & $(\mathbf{w}_1,4)$ & \textcolor{red}{$(\mathbf{w}_1,5)$} & $(\mathbf{w}_1,4)$ & $(\mathbf{w}_1,3)$ & $(\mathbf{w}_1,2)$ & \textcolor{red}{$(\mathbf{w}_1, 3)$} & \textcolor{red}{$(\mathbf{w}_1,4)$} & \textcolor{red}{$(\mathbf{w}_1,3)$} & \textcolor{red}{$(\mathbf{w}_1,2)$}\\
		input & ---  & $0$ & $1$ & $1$ & $0$ & $0$ & $\textcolor{red}{1}$ & $0$& $0$ & $0$ &$1$ &$1$ &$0$ &$0$ \\
		output &  --- & $\mathbf{w}_2$ & $\mathbf{w}_1$ & $\mathbf{w}_1$ & $\mathbf{w}_2$ & $\mathbf{w}_2$ & \textcolor{red}{$\mathbf{w}_1$} & $\mathbf{w}_2$& $\mathbf{w}_2$ & $\mathbf{w}_2$ &\textcolor{red}{$\mathbf{w}_1$} &\textcolor{red}{$\mathbf{w}_1$} &\textcolor{red}{$\mathbf{w}_2$} &\textcolor{red}{$\mathbf{w}_2$}\\\bottomrule
	\end{tabular}
}
	\captionof{table}{Encoding and decoding processes in Fig.~\ref{fig::errormodel1}.} \label{tab::errormodel1}
\end{table}

\begin{table}
	\centering
	\scalebox{0.68}{
		\begin{tabular}{ccccccccccccccc}
			\toprule
			$t$&  0 & 1 & 2 & 3 & 4 & 5 & 6 & 7 & 8 & 9 & 10 & 11 & 12 & 13    \\\midrule
			$(s^e, N^e)$ & $(\mathbf{w}_1, 5)$ & $(\mathbf{w}_1, 4)$ & $(\mathbf{w}_1,5)$ &  $(\mathbf{w}_1,6)$  & $(\mathbf{w}_1,5)$  & $(\mathbf{w}_1,4)$ & $(\mathbf{w}_1,3)$ & $(\mathbf{w}_1,2)$ & $(\mathbf{w}_1,3)$ & $(\mathbf{w}_1,4)$ & $(\mathbf{w}_1, 5)$ & $(\mathbf{w}_1,4)$ & $(\mathbf{w}_1,5)$ & $(\mathbf{w}_1,4)$\\
			input & ---  & $\mathbf{w}_2$ & $\mathbf{w}_1$ & $\mathbf{w}_1$ & $\mathbf{w}_2$ & $\mathbf{w}_2$ & $\mathbf{w}_2$ & $\mathbf{w}_2$& $\mathbf{w}_1$ & $\mathbf{w}_1$ &$\mathbf{w}_1$ &$\mathbf{w}_2$ &$\mathbf{w}_1$ &$\mathbf{w}_2$ \\
			output & ---  & $0$ & $1$ & $1$ & $0$ & $0$ & $0$ & $0$& $1$ & $1$ &$1$ &$0$ &$1$ &$0$\\\midrule
			$(s^d, N^d)$ &$(\mathbf{w}_1, 5)$ & $(\mathbf{w}_1, 4)$ & $(\mathbf{w}_1,5)$ &  \textcolor{red}{$(\mathbf{w}_1,4)$}  & $(\mathbf{w}_1,3)$  & $(\mathbf{w}_1,2)$ & $(\mathbf{w}_1,1)$ & \textcolor{red}{$(\mathbf{w}_2,0)$} & \textcolor{red}{$(\mathbf{w}_2,-1)$} & \textcolor{red}{$(\mathbf{w}_2,-2)$} & \textcolor{red}{$(\mathbf{w}_2, -3)$} & \textcolor{red}{$(\mathbf{w}_2,-2)$} & \textcolor{red}{$(\mathbf{w}_1,-3)$} & \textcolor{red}{$(\mathbf{w}_1,-2)$}\\
			input & ---  & $0$ & $1$ & $\textcolor{red}{0}$ & $0$ & $0$ & $0$ & $0$& $1$ & $1$ &$1$ &$0$ &$1$ &$0$\\
			output & ---  & $\mathbf{w}_2$ & $\mathbf{w}_1$ & \textcolor{red}{$\mathbf{w}_2$} & $\mathbf{w}_2$ & $\mathbf{w}_2$ & $\mathbf{w}_2$ & $\mathbf{w}_2$& \textcolor{red}{$\mathbf{w}_2$}  & \textcolor{red}{$\mathbf{w}_2$}  &\textcolor{red}{$\mathbf{w}_2$}  &\textcolor{red}{$\mathbf{w}_1$}  &\textcolor{red}{$\mathbf{w}_2$}  &\textcolor{red}{$\mathbf{w}_1$} \\\bottomrule
		\end{tabular}
	}
	\captionof{table}{Encoding and decoding processes in Fig.~\ref{fig::errormodel2}.} \label{tab::errormodel2}
\end{table}

In order to analyze the error propagation behavior of direct shaping codes, we make use of a result about one-dimensional random walks.
Here we use the encoding process as an example. For a two-word dictionary  with input words $\mathbf{w}_1$ and $\mathbf{w}_2$ and respective counts $n_1^e(t)$ and $n_2^e(t)$, suppose $n_1^e(t_0)\neq n_2^e(t_0)$ at time $t_0$. We say a \textit{recurrence} occurs if, at some future time $t>t_0$, $n_{1}^e(t)= n_2^e(t)$. The following theorem, stated in~\cite[Theorem 4.8.9]{Durrett}, determines the recurrence probability.

\begin{theorem}\label{thm::onedimrecurrence}
	Consider two input words $\mathbf{w}_1,  \mathbf{w}_2$, with probabilities $P_1, P_2$ respectively ($P_1+P_2=1$, $P_1 \geq P_2$). Let $N^e(t)=n_1^e(t)-n_2^e(t)$. At time $t_0$, the probability $Q$ that a recurrence involving $\mathbf{w}_1,  \mathbf{w}_2$ will occur in the future is
	
	\begin{equation}
	Q=\begin{cases}\left(\frac{P_2}{P_1}\right)^{N^e(t_0)} \quad \mbox{if $N^e(t_0)>0$,}\\1\quad\quad\quad\,\,\, \quad\,\,\mbox{if $N^e(t_0)\leq 0$.}
	\end{cases} 
	\end{equation}\qed
\end{theorem}

\begin{figure}
	\centering
	\includegraphics[width=0.6\columnwidth]{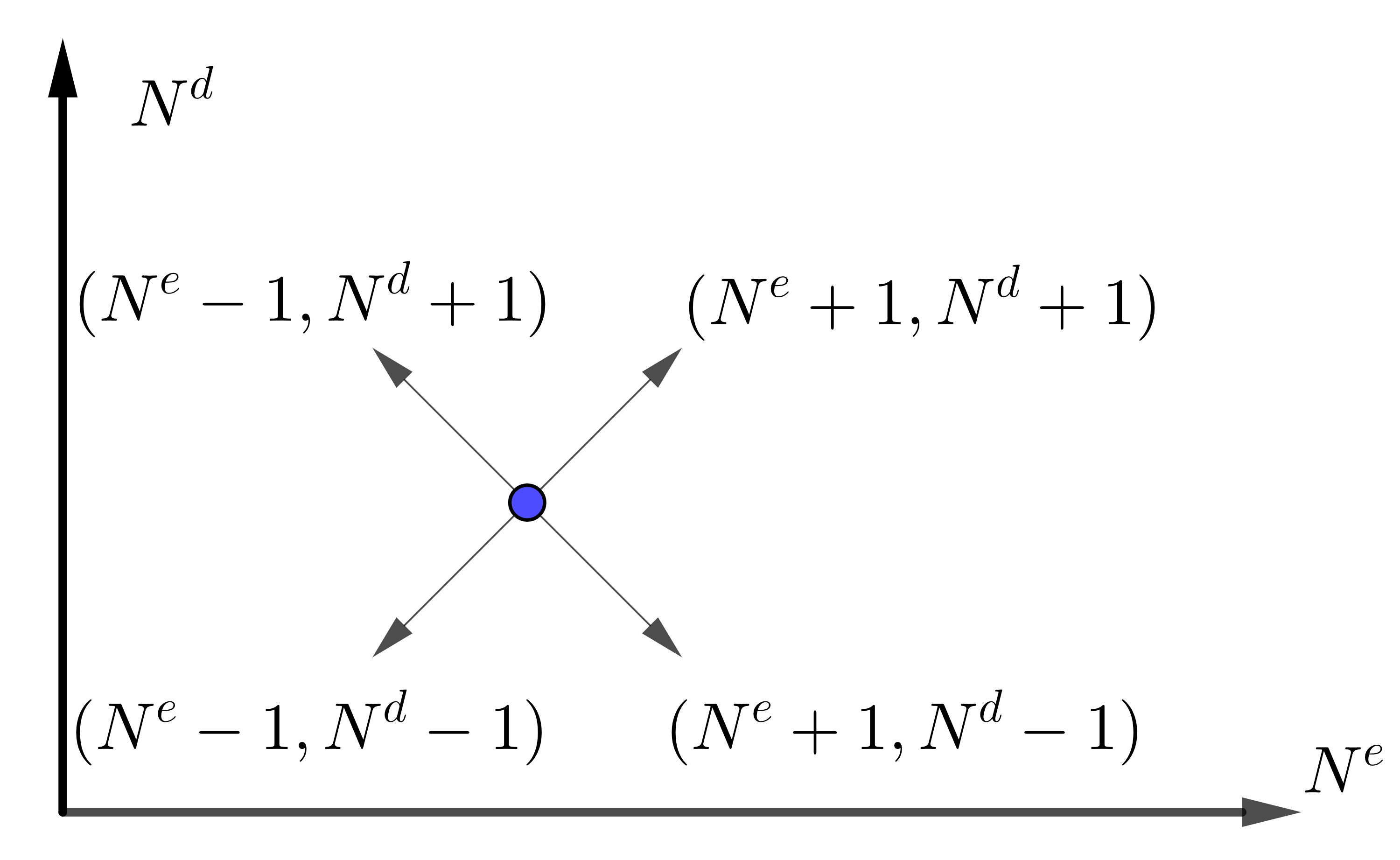}
	\caption{Two dimensional random walk on $(N_i^e,N_i^d)$-plane}
	\label{fig::twodimrw}
\end{figure}

Here we assume that the SLC flash memory is a binary symmetric channel with transition probability $\rho$. The pair $\{N^e(t),N^d(t)\}$ acts like a two-dimensional random walk, as shown in Fig.~\ref{fig::twodimrw}. The transition probabilities of this random walk are
\begin{equation}
\begin{small}
\begin{aligned}
&P(\{N^e(t), N^d(t)\}\rightarrow \{N^e(t)+1,N^d(t)+1\})=(1-\rho)P_1,\\
&P(\{N^e(t), N^d(t)\}\rightarrow \{N^e(t)+1,N^d(t)-1\})=\rho P_1,\\
&P(\{N^e(t), N^d(t)\}\rightarrow \{N^e(t)-1,N^d(t)-1\})=(1-\rho)P_2,\\
&P(\{N^e(t), N^d(t)\}\rightarrow \{N^e(t)-1,N^d(t)+1\})=\rho P_2.
\end{aligned}
\end{small}
\end{equation}

We say a recurrence occurs if, at future time $t>t_0$, either $N^e(t)$ or $N^d(t)$ equals to zero. We denote by $E^e$ the event that the recurrence occurs during the encoding process (at future time $t>t_0$, $N^e(t)$ equals to zero) and $E^d$ the event that the recurrence occurs during the decoding process. The recurrence probability is
\begin{equation}
\label{func::two_words_recurrence_def}
Q= \text{Pr}\{E^e\cup E^d\}.
\end{equation}
Using Theorem~\ref{thm::onedimrecurrence}, we can calculate $\text{Pr}\{E^e\}$ and $\text{Pr}\{E^d\}$. When $\rho \geq 0.5$, $N^d(t)$ increases by 1 with probability 
\begin{equation}
\begin{aligned}
	P_d& = (1-\rho)P_1 +\rho P_2 \\
	& = (1-\rho) P_1 +\rho (1-P_1)\\
	& = \frac{1}{2}-\frac{1}{2}(1-2\rho)(1-2P_1) \leq \frac{1}{2}.
\end{aligned}
\end{equation}
This indicates that $\text{Pr}\{E^d\} = 1$ when $N^d(t) >0$. Similarly, when $N^e(t) < 0$, from Theorem~\ref{thm::onedimrecurrence} we have $\text{Pr}\{E^e\} = 1$. To avoid these situations, we assume that $\rho < 0.5$ and $N^e(t_0),N^d(t_0) >0$ at time $t_0$. 
we say the pair $\{N^e(t_0),N^d(t_0)\}$ is \emph{stable} if $N^e(t_0),N^d(t_0) >0$. 
The following theorem provides an upper bound on $Q$.

\begin{theorem}
	\label{thm::twodimupperbound}
	For a two-word dictionary with a stable pair $\{N^e(t_0),N^d(t_0)\}$ at time $t_0$, the probability $Q$ that a recurrence will occur in the future is bounded by
	\begin{equation}
	  Q \leq  \left(\frac{P_2}{P_1}\right)^{N^e(t_0)} + \left[\frac{\rho P_1 + (1-\rho)P_2}{(1-\rho) P_1 + \rho P_2}\right]^{N^d(t_0)}.
	\end{equation}
\end{theorem} 

\begin{IEEEproof}
		A simple union bound yields
		\begin{equation}
		\label{func::two_word_upper_bound}
		\begin{aligned}
		Q &= \text{Pr}\{E^e\cup E^d\}  \leq\text{Pr}\{E^e\}+ \text{Pr}\{E^d\}.
		\end{aligned}
				\end{equation}
		From Theorem~\ref{thm::onedimrecurrence}, we have
		\begin{equation}
		\label{func::recurrence_encoder}
		\text{Pr}\{E^e\} = \left(\frac{P_2}{P_1}\right)^{N^e(t_0)}
		\end{equation}
		and 
			\begin{equation}
			\label{func::recurrence_decoder}
		\text{Pr}\{E^d\} = \left[\frac{\rho P_1 + (1-\rho)P_2}{(1-\rho) P_1 + \rho P_2}\right]^{N^d(t_0)}.
		\end{equation}
		Combining equations~(\ref{func::two_word_upper_bound}),~ (\ref{func::recurrence_encoder}) and~(\ref{func::recurrence_decoder}) completes the proof.
\end{IEEEproof}

\subsection{Recurrence probability of a pair of words in a general dictionary}
In this subsection, we study the recurrence probability of a pair of words in the general dictionary with $2^m$ entries.
We first establish some notation. Assume  $t$ data words have been encoded. 
Let $\mathbf{n}^e(t)=\{n^e_1(t),n^e_2(t),n^e_3(t),\ldots,n^e_{2^m}(t)\}$ be the word counts and let $N^e_i(t)=n^e_i(t)-n^e_{i+1}(t)$ denote the \textit{distance} between the $i$th and $(i+1)$st words. Similarly, for the decoding process, let $\mathbf{n}^d(t)=\{n^d_1(t),n^d_2(t),\ldots,n^d_{2^m}(t)\}$ be the word counts and let $N^d_i(t)=n^d_i(t)-n^d_{i+1}(t)$ denote the distance. We say the pair $\{\mathbf{n}^e(t),\mathbf{n}^d(t)\}$ is \textit{stable} if $n^e_1(t)>n^e_2(t)>\ldots>n^e_{2^m}(t)$ and $n^d_1(t)>n^d_2(t)>\ldots>n^d_{2^m}(t)$. We sometimes use the term stable to describe a dictionary that has a stable $\{\mathbf{n}^e(t),\mathbf{n}^d(t)\}$. 
When the dictionary is stable, the probability of reading word $\mathbf{w_i}$ from flash memory is 
\begin{equation}
	\label{func::decodingtransitionprob}
	P_i^d = \sum_j \rho^{\ham{i,j}} (1-\rho)^{m-\ham{i,j}}P_j,
\end{equation}
where $\ham{i,j}$ is the Hamming distance between $\mathbf{w_i}$ and $\mathbf{w_j}$. Here we always assume that $\rho$ is small enough so that $P_1^d \geq P_2^d \geq \ldots \geq P_{2^m}^d$.

We assume that the dictionary is stable at some $t_0$. 
We denote by $\{i,i+1\}^e$ the event that a recurrence occurs between the $i$th and $(i+1)$st words during the encoding process and by $\{\oline{i,i+1}\}^e$ its complement. Similarly, we denote by $\{i,i+1\}^d$ the event that a recurrence occurs between the $i$th and $(i+1)$st words during the decoding process and by $\{\oline{i,i+1}\}^d$ its complement. In this subsection, we study the recurrence probability $Q_i(N_i^e(t), N_i^d(t)) = \text{Pr}\{\{{i,i+1}\}^e \bigcup \{{i,i+1}\}^d\}$. We first note the following lemma which is a direct corollary of Theorem~\ref{thm::twodimupperbound}.
\begin{lemma}
	\label{lemma::twowordrecurrence}
		For a general dictionary with a stable pair $\{N_i^e(t_0), N_i^d(t_0)\}$  at time $t_0$, the probability $Q_i(N_i^e(t_0), N_i^d(t_0)) $ that a recurrence will occur in the future between $\mathbf{w}_i$ and $\mathbf{w}_{i+1}$ is bounded by 
		\begin{equation}
		Q_i(N_i^e(t_0), N_i^d(t_0)) \leq \left(\frac{P_{i+1}}{P_i}\right)^{N_i^e(t_0)} + \left(\frac{P_{i+1}^d}{P_i^d}\right)^{N^d(t_0)},
		\end{equation}
		where $\{P_i^d\}$ is given by equation~(\ref{func::decodingtransitionprob}).\qed
\end{lemma}

\begin{figure}
	\centering
	\includegraphics[width=0.6\columnwidth]{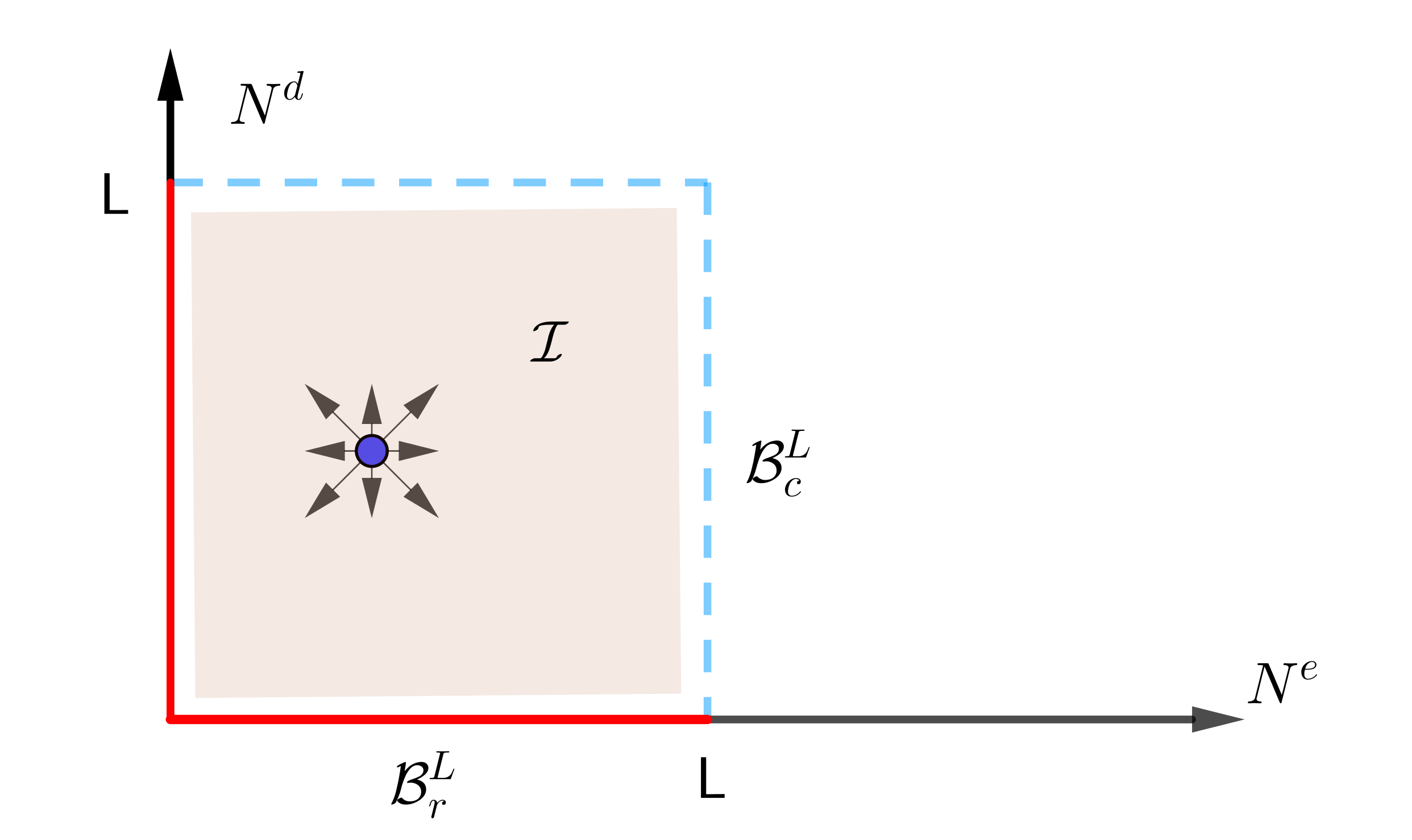}
	\caption{Two dimensional random walk in a general dictionary}
	\label{fig::dicttwodimrw}
\end{figure}

Here we provide a way to numerically approximate a lower bound of $Q_i(N_i^e(t_0), N_i^d(t_0)) $. Consider a closed region in the $(N_i^e,N_i^d)$-plane defined by four line segments, which are divided into two groups $\mathcal{B}^L_r$ and $\mathcal{B}^L_c$.
\begin{equation}
\begin{aligned}
	&\mathcal{B}^L_r:  N_i^e = 0, \,0\leq N_i^d\leq L,\,\text{and} \, N_i^d = 0 ,\, 0\leq N_i^e \leq L, \\
	&\mathcal{B}^L_c:  N_i^e = L, \,0<N_i^d\leq L,\,\text{and} \, N_i^d = L ,\, 0< N_i^e \leq L ,
\end{aligned}
\end{equation}
where $L > \max\{N_i^e(t_0), N_i^d(t_0)\}$. As shown in Fig.~\ref{fig::dicttwodimrw}, $\mathcal{B}^L_r$ is represented by the red line and $\mathcal{B}^L_c$ is represented by the blue dashed line. The interior region is denoted by $\mathcal{I}$.

Unlike the two-word dictionary, where the pair $(N_i^e, N_i^d)$  behaves like a two-dimensional random walk with only 4 directions of movement, the pair $(N_i^e, N_i^d)$ in a general dictionary can move in 8 directions or stand still. Denote the probability of going from $(N_i^e, N_i^d)$ to $(N_i^e+x,N_i^d+y)$ by $P^i_{\{x,y\}}$, $x,y\in\{-1,0,1\}$. The transition probabilities of this random walk are shown in Table~\ref{table::twodim}.
Note that in the first four entries of the table, the summations actually include only one term. They have been written as summations to highlight the similar form of the various transition probabilities. 
\begin{table}[h!]
	\centering
	\begin{normalsize}
		\begin{tabular}{@{}cl@{}}
			\toprule
			Direction&  \multicolumn{1}{c}{Transition Probability} \\ \midrule
			$\{1,1\}$&$P^i_{\{1,1\}}=\sum_{j = i} \rho^{\ham{i,j}}(1-\rho)^{m-\ham{i,j}} P_i$\\
			$\{-1,-1\}$&$P^i_{\{-1,-1\}}=\sum_{j = i+1} \rho^{\ham{i+1,j}}(1-\rho)^{m-\ham{i+1,j}} P_{i+1}$\\
			$\{1,-1\}$&$P^i_{\{1,-1\}}=\sum_{j = i+1} \rho^{\ham{i,j}}(1-\rho)^{m-\ham{i,j}} P_i$\\
			$\{-1,1\}$&$P^i_{\{-1,1\}}=\sum_{j = i} \rho^{\ham{i+1,j}}(1-\rho)^{m-\ham{i+1,j}} P_{i+1}$\\
			$\{1,0\}$&$P^i_{\{1,0\}}=\sum_{j \neq i, i+1} \rho^{\ham{i,j}}(1-\rho)^{m-\ham{i,j}} P_i$\\
			$\{-1,0\}$&$P^i_{\{-1,0\}}=\sum_{j \neq i, i+1} \rho^{\ham{i+1,j}}(1-\rho)^{m-\ham{i+1,j}} P_{i+1}$\\
			$\{0,1\}$&$P^i_{\{0,1\}}=\sum_{j \neq i, i+1} \rho^{\ham{i,j}}(1-\rho)^{m-\ham{i,j}} P_j$\\
			$\{0,-1\}$&$P^i_{\{0,-1\}}=\sum_{j \neq i, i+1} \rho^{\ham{i+1,j}}(1-\rho)^{m-\ham{i+1,j}} P_j$\\
			$\{0,0\}$&$P^i_{\{0,0\}}=1 - \sum_{x, y \in \{-1,0,1\}, \{x,y\}\neq \{0,0\} }P^i_{\{x,y\}}$\\
			\\\bottomrule
		\end{tabular}    
	\end{normalsize}   
	\caption{Transition probability of a random walk in a general dictionary.}
	\label{table::twodim}
\end{table}

For any point $(N^e, N^d)$ in the closed region, denote by $E_L(N^e, N^d)$ (or $E_L$ for convenience) the event that a random walk starting at  $(N^e, N^d)$ will hit boundary group $\mathcal{B}^L_r$ before it hits $\mathcal{B}^L_c$  and denote by $\widetilde{Q_i}(E_L)$ its probability. Clearly $E_{L_2} \subseteq E_{L_1}$ when $L_2> L_1$, because if a random walk hits $\mathcal{B}^{L_1}_r$ before hitting $\mathcal{B}^{L_1}_c$, it also hits $\mathcal{B}^{L_2}_r$ before hitting $\mathcal{B}^{L_2}_c$. Therefore we have 
\begin{equation}
\begin{aligned}
&\widetilde{Q}_i(E_L)\leq \widetilde{Q}_i(E_{L+1})\leq \ldots \leq Q_i , \\
&\Rightarrow \widetilde{Q}_i(N^e, N^d) \overset{\rm{def}}{=} \limsup_{L\rightarrow \infty} \widetilde{Q}_i(E_L) \leq Q_i(N^e, N^d).
\end{aligned}
\end{equation}
$\widetilde{Q}_i(E_L(N^e, N^d))$ has the following properties.
\begin{equation}
\label{func::transitionproperty1}
\widetilde{Q}_i(E_L(0, N^d))= \widetilde{Q}_i(E_L(N^e,0)) = 1,
\end{equation}
\begin{equation}
\label{func::transitionproperty2}
	\begin{aligned}
		\widetilde{Q}_i(E_L(L, N^d)) =0 \text{ when } N^d >0, \\
		\widetilde{Q}_i(E_L(N^e, L)) =0\text{ when } N^e >0,
	\end{aligned}
\end{equation}
\begin{equation}
\label{func::transitionproperty3}
\widetilde{Q}_i(E_L(N^e,N^d))= \frac{\sum_{\{x,y\}\neq \{0,0\} }P^i_{\{x,y\}}\widetilde{Q}_i(E_L(N^e + x, N^d +y))}{1-P^i_{\{0,0\}}}.
\end{equation}

To find $\widetilde{Q}_i(E_L(N^e_i(t_0), N^d_i(t_0)))$, we first define the recurrence probability vector $\mathbf{Q}_i^L$ by 
\begin{equation}
\mathbf{Q}_i^L =[\underbrace{\widetilde{Q}_i(E_L(0,0)), \ldots}_{\mathcal{B}^L_r}, \underbrace{\widetilde{Q}_i(E_L(1,L)),\ldots}_{\mathcal{B}^L_c},\underbrace{\widetilde{Q}_i(E_L(1,1))\ldots}_{\mathcal{I}}]^\top
\end{equation}
where, within each subvector, the ordering of the values corresponds to the lexicographical ordering of the locations $(N^e_i(t_0), N^d_i(t_0))$.
Updating $\mathbf{Q}_i^L$ based on equations~(\ref{func::transitionproperty1}),~(\ref{func::transitionproperty2}) and~(\ref{func::transitionproperty3}) can then be represented by a matrix multiplication $\mathbf{T}\mathbf{Q}_i^L$.
The transition matrix $\mathbf{T}$ has the form
\begin{equation}
\mathbf{T} = \left[\begin{array}{c:c}
\mathbf{I}&\mathbf{0}\\\hdashline[3pt/3pt]\mathbf{R}&\mathbf{S}\end{array}\right].
\end{equation}
It is easy to check that $\mathbf{T}$ satisfies the following properties.
\begin{itemize}
	\item $\mathbf{T}$ is non-negative.
	\item The row sum $\sum_j t_{ij} = 1$ for all $i$.
	\item $\mathbf{I}$ is an identity matrix of size $4L$, representing the boundary conditions.
	\item $\mathbf{S}$ is irreducible and there exists row $i$ such that $\sum_j s_{ij} < 1$.
	\item $\mathbf{T}^h$ can be represented as 
	\begin{equation}
	\label{function::poweroft}
\mathbf{T}^h =  \left[\begin{array}{c:c}
\mathbf{I}&\mathbf{0}\\\hdashline[3pt/3pt]\mathbf{R}(\mathbf{I} + \mathbf{S} +\ldots + \mathbf{S}^{h-1} )&\mathbf{S}^h\end{array}\right].
	\end{equation}
\end{itemize}
Since $\mathbf{S}$ has rows with row sum strictly less than 1, we can always find an irreducible matrix $\widetilde{\mathbf{S}}$ such that $\widetilde{\mathbf{S}} \geq \mathbf{S}$, $\widetilde{\mathbf{S}} \neq \mathbf{S}$ and its row sum $\sum_j \widetilde{s}_{ij} = 1$ for all rows. From~\cite[Lemma 1.2]{Nussbaum}, the Perron root $\lambda(\mathbf{S})$ of $\mathbf{S}$ is strictly less than that of $\widetilde{\mathbf{S}}$ and
\begin{equation}
\lambda(\mathbf{S}) < \lambda(\widetilde{\mathbf{S}}) = 1.
\end{equation}
Thus we have 
\begin{equation}
\label{function::slimit}
\lim_{h\rightarrow \infty} \mathbf{S}^h = 0,
\end{equation}
\begin{equation}
\label{function::sserieslimit}
\lim_{h\rightarrow \infty} \mathbf{I} + \mathbf{S} +\ldots + \mathbf{S}^{h-1} = (\mathbf{I}-\mathbf{S})^{-1}.
\end{equation}
By combining equations~(\ref{function::poweroft}),~(\ref{function::slimit}) and~(\ref{function::sserieslimit}), we have 
\begin{equation}
\label{function::tlimit}
\begin{aligned}
\lim_{h\rightarrow \infty} \mathbf{T}^h & = \lim_{h\rightarrow \infty}  \left[\begin{array}{c:c}
\mathbf{I}&\mathbf{0}\\\hdashline[3pt/3pt]\mathbf{R}(\mathbf{I} + \mathbf{S} +\ldots + \mathbf{S}^{h-1} )&\mathbf{S}^h\end{array}\right]=  \left[\begin{array}{c:c}
\mathbf{I}&\mathbf{0}\\\hdashline[3pt/3pt]\mathbf{R}(\mathbf{I}-\mathbf{S})^{-1} &\mathbf{0}\end{array}\right].
\end{aligned}
\end{equation}
Equation~(\ref{function::tlimit}) indicates that $\widetilde{Q}_i(E_L(N^e,N^d))$ is determined solely by the boundary values (equations~(\ref{func::transitionproperty1}) and~(\ref{func::transitionproperty2})) and can be numerically approached by Algorithm~\ref{algo::calculateq}. By increasing $L$, $\widetilde{Q}_i(E_L(N^e,N^d))$ can be used to approach $\widetilde{Q}_i(N^e, N^d)$, which is the lower bound on $Q_i(N^e, N^d)$.
 \begin{algorithm}[ht]
 	\caption{Approximating $\widetilde{Q}_i(E_L(N_i^e(t_0),N_i^d(t_0)))$}
 	\label{algo::calculateq}
 	\algsetup{
 		linenosize=\small
 	}
 	\begin{algorithmic}[1]
 		\REQUIRE Area size $L$, index $(N_i^e(t_0),N_i^d(t_0))$, maximum iteration number $J$.
 		\ENSURE Approximation of $\widetilde{Q}_i(E_L(N_i^e(t_0),N_i^d(t_0)))$.
 		\STATE Initialize $\mathbf{Q}_i^L$ such that $\widetilde{Q}_i(E_L(0,N^d))= \widetilde{Q}_i(E_L(N^e,0)) = 1$ and $\widetilde{Q}_i(E_L(N^e,L)) = \widetilde{Q}_i(E_L(L,N^d)) = 0$. The remaining $\widetilde{Q}_i(E_LN^e,N^d))$ are set to 0. 
 		\FOR{each iteration $j=1, \ldots J$}
 		\STATE Update $\widetilde{Q}_i(E_L(N^e,N^d))$ based on equations~(\ref{func::transitionproperty1}),~(\ref{func::transitionproperty2}) and~(\ref{func::transitionproperty3}).
 		\ENDFOR
 	\end{algorithmic}
\end{algorithm}

\begin{figure*}[t]
	\centering
	\begin{subfigure}[t]{0.32\textwidth}
		\includegraphics[width=\linewidth]{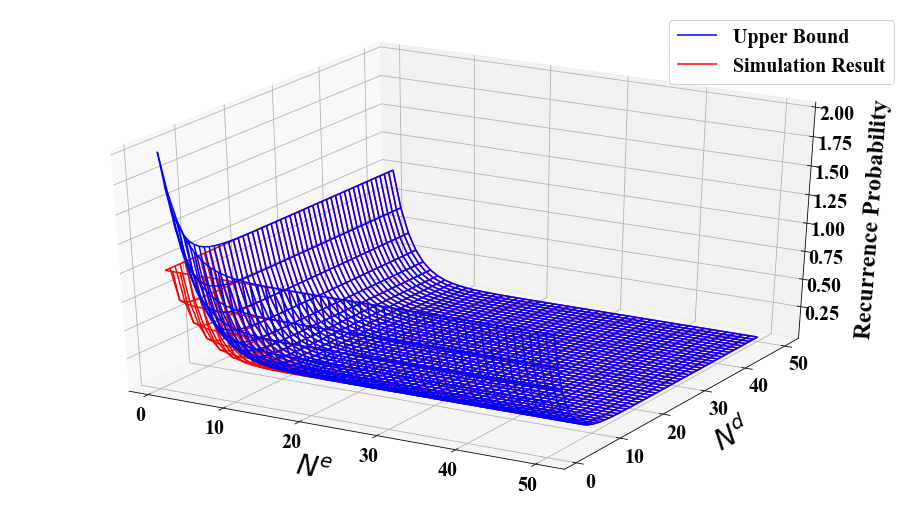}
		\caption{$Q_i(N^e,N^d)$} \label{fig::upperboundandsimulation}
	\end{subfigure}
	\hspace*{\fill} 
	\begin{subfigure}[t]{0.32\textwidth}
		\includegraphics[width=\linewidth]{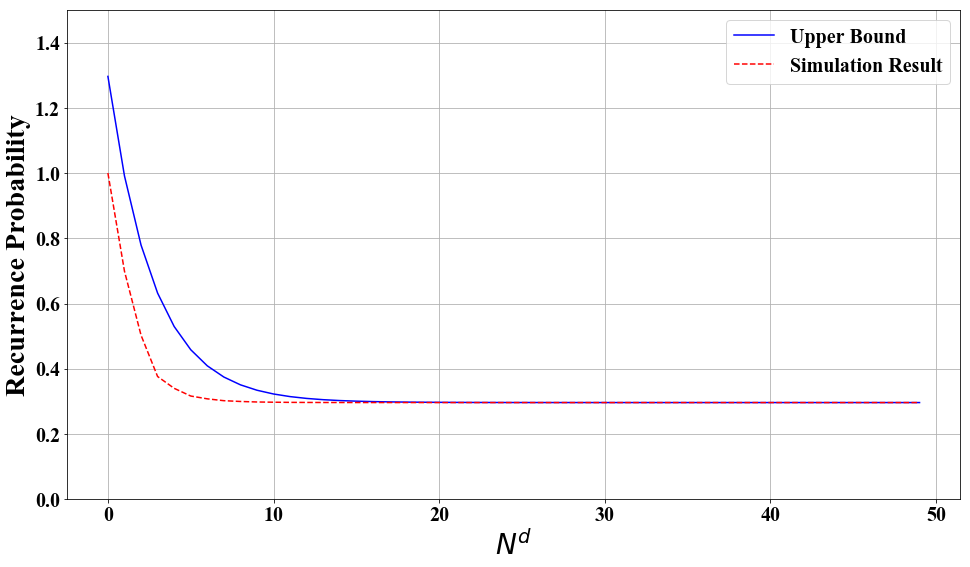}
		\caption{$N^e = 3$.} \label{fig::nefix}
	\end{subfigure}
	\hspace*{\fill} 
	\begin{subfigure}[t]{0.32\textwidth}
		\includegraphics[width=\linewidth]{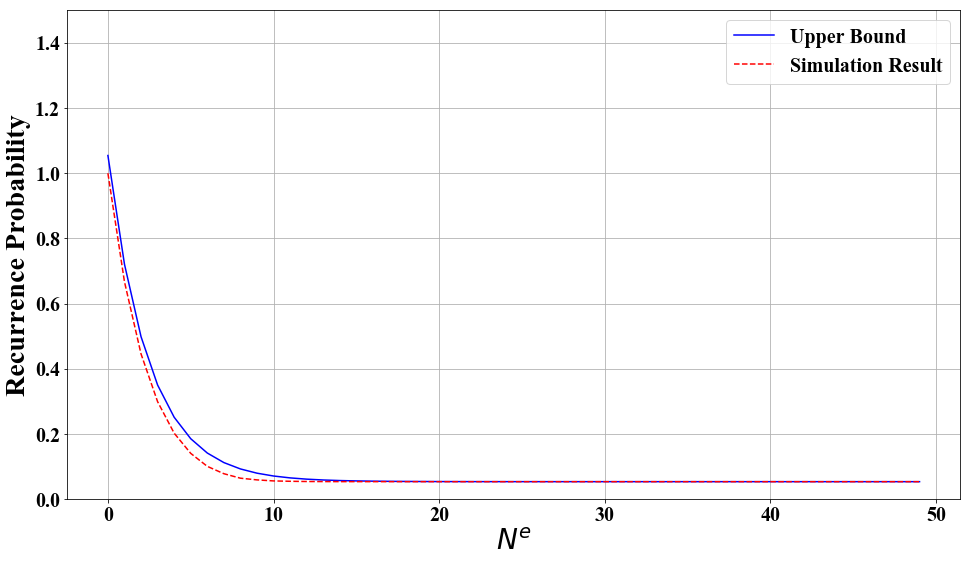}
		\caption{$N^d = 8$} \label{fig::ndfix}
	\end{subfigure}
	\bigskip
	\caption{Recurrence probabilities of a two-word dictionary with distribution $P=\{0.6,0.4\}$ and transition probability $\rho = 0.05$.}
	\label{fig::recurrenceprobabilityofatwoworddictionary}
\end{figure*}

\begin{remark}
	This method can also be applied to the two-word dictionary. However, since the random walk moves in only four directions, as shown in Fig.~\ref{fig::twodimrw}, point $(N^e,N^d)$ is only connected to other points $(\widetilde{N}^e,\widetilde{N}^d)$  when $N^e + N^d$ and  $\widetilde{N}^e + \widetilde{N}^d$ are both even or odd. This implies that $\mathbf{S}$ is not irreducible. To solve this problem, we can further divide $\mathbf{S}$ into two irreducible components, based on the parity of $N^e+ N^d$. The transition matrix $\mathbf{T}$ can then be represented as 
	\begin{equation}
	\mathbf{T} = \left[\begin{array}{c:cc}
	\mathbf{I}&\mathbf{0}&\mathbf{0}\\\hdashline[3pt/3pt]\mathbf{R}_1&\mathbf{S}_1 &\mathbf{0}\\\mathbf{R}_2&\mathbf{0}&\mathbf{S}_2\end{array}\right]
	\end{equation}
	and
	\begin{equation}
	\lim_{h\rightarrow \infty}\mathbf{T}^h = \left[\begin{array}{c:cc}
	\mathbf{I}&\mathbf{0}&\mathbf{0}\\\hdashline[3pt/3pt]\mathbf{R}_1(\mathbf{I}-\mathbf{S}_1)^{-1}&\mathbf{0} &\mathbf{0}\\\mathbf{R}_2(\mathbf{I}-\mathbf{S}_2)^{-1}&\mathbf{0}&\mathbf{0}\end{array}\right].
	\end{equation}
	In Fig.~\ref{fig::recurrenceprobabilityofatwoworddictionary}, the numerical results on the  recurrence probabilities of a two-word dictionary with distribution $P = \{0.6, 0.4\}$ and transition probability $\rho = 0.05$ are presented. Both the upper bound in Theorem~\ref{thm::twodimupperbound} and the simulation result using Algorithm~\ref{algo::calculateq} are shown. We can see that when $N^e$ or $N^d$ is large, the upper bound is tight.
	
\end{remark}
\subsection{Recurrence probability of a general dictionary}
In this subsection we discuss the recurrence probability of the whole dictionary. Let $W$ be the event that \textit{a recurrence occurs involving any two \textit{adjacent} words in a stable dictionary}, we have
\begin{equation}
W = \left(\bigcup_{i=1}^{2^{m-1}-1}\{i,i+1\}^e\right)\cup \left(\bigcup_{i=1}^{2^{m-1}-1}\{i,i+1\}^d\right).
\end{equation}
Since a recurrence between words $\mathbf{w}_i$ and $\mathbf{w}_j$ always includes recurrence between adjacent words, we know that the probability of a recurrence involving any two words in the dictionary, denoted $P_W$, is
\begin{equation}
P_W=\text{Pr}\{W\}.
\end{equation}
Notice that $\{i,i+1\}^e$ and $\{i+1,i+2\}^e$ are not independent, but $\{i,i+1\}^e$ and $\{i+2,i+3\}^e$ are independent. This is also true for the decoding process. We now derive an upper bound on $P_W$.  We have the following lemma.

\begin{lemma}
	\label{lemma::generalupperbound}
	The probability of recurrence $P_W$ satisfies
	\begin{equation}
	\small
	\label{func::generalupperbound}
	\begin{split}
		P_W\leq &4-\prod_{i=1}^{2^{m-1}}\left(1-\frac{P_{2i}}{P_{2i-1}}^{N^e_{2i-1}(t)}\right)-\prod_{i=1}^{2^{m-1}-1}\left(1-\frac{P_{2i+1}}{P_{2i}}^{N^e_{2i}(t)}\right)-\prod_{i=1}^{2^{m-1}}\left(1-\frac{P^d_{2i}}{P^d_{2i-1}}^{N^d_{2i-1}(t)}\right)-\prod_{i=1}^{2^{m-1}-1}\left(1-\frac{P^d_{2i+1}}{P^d_{2i}}^{N_{2i}^d(t)}\right).
		\end{split}
	\end{equation}
\end{lemma}

\begin{IEEEproof}
See Appendix~\ref{appen::generalupperbound}.
\end{IEEEproof}
Here we assume that the first $t$ data words are encoded and decoded correctly, i.e. $\mathbf{n}^e(t)=\mathbf{n}^d(t)$.
We denote by $A(\mathbf{n}^e(t))$ the right side of equation (\ref{func::generalupperbound}).  Let $P(\mathbf{n}^e(t))$ denote the probability that after $t$ steps the word counts are $\mathbf{n}^e(t)$. Then, we have

\begin{equation}
	P(\mathbf{n}^e(t))=\binom{t}{n^e_1(t),\ldots,n^e_{2^m}(t)}P_1^{n^e_1(t)}\ldots P_{2^m}^{n^e_{2^m}(t)}.
\end{equation}
By combining Lemma \ref{lemma::generalupperbound} and the law of total probability, we have the following theorem.
\begin{theorem}
	\label{thm::upperboundoftotalprobability}
After $t$ data words are encoded and decoded correctly, the probability $P(t)$ that the dictionary will be unstable satisfies
\begin{equation}
	P(t)\leq \sum_{\substack{ \mathbf{n}^e(t)\\ \mbox{stable}}}A(\mathbf{n}^e(t))P(\mathbf{n}^e(t))+\sum_{\substack{\mathbf{n}^e(t)\mbox{ not}\\ \mbox{ stable}}}P(\mathbf{n}^e(t)).
\end{equation}\qed
\end{theorem}

\begin{figure*}[t]
	\centering
	\begin{subfigure}[t]{0.32\textwidth}
		\includegraphics[width=\linewidth]{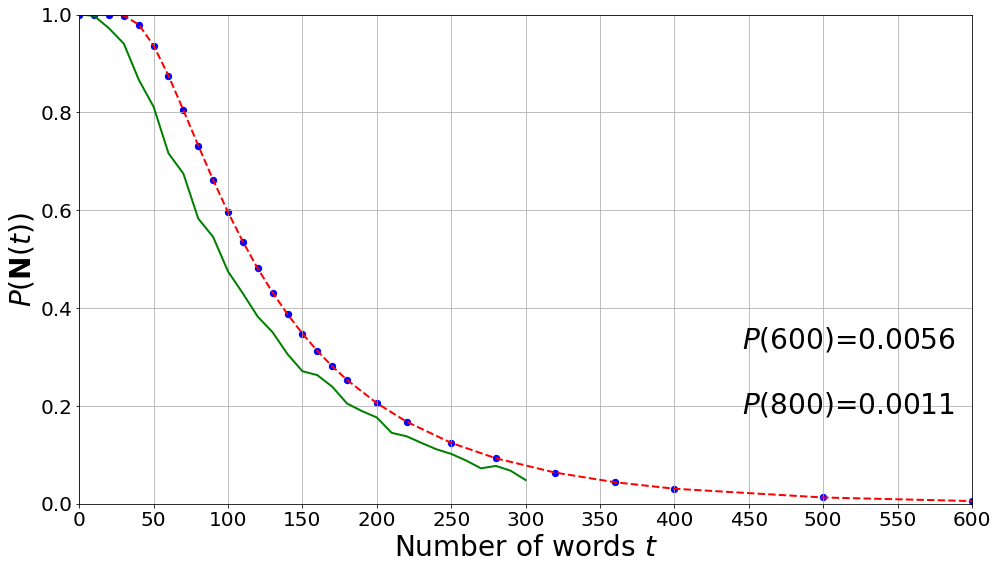}
		\caption{\textit{$\rho = 0.05$}. } \label{fig::rho_05}
	\end{subfigure}
	\hspace*{\fill} 
	\begin{subfigure}[t]{0.32\textwidth}
		\includegraphics[width=\linewidth]{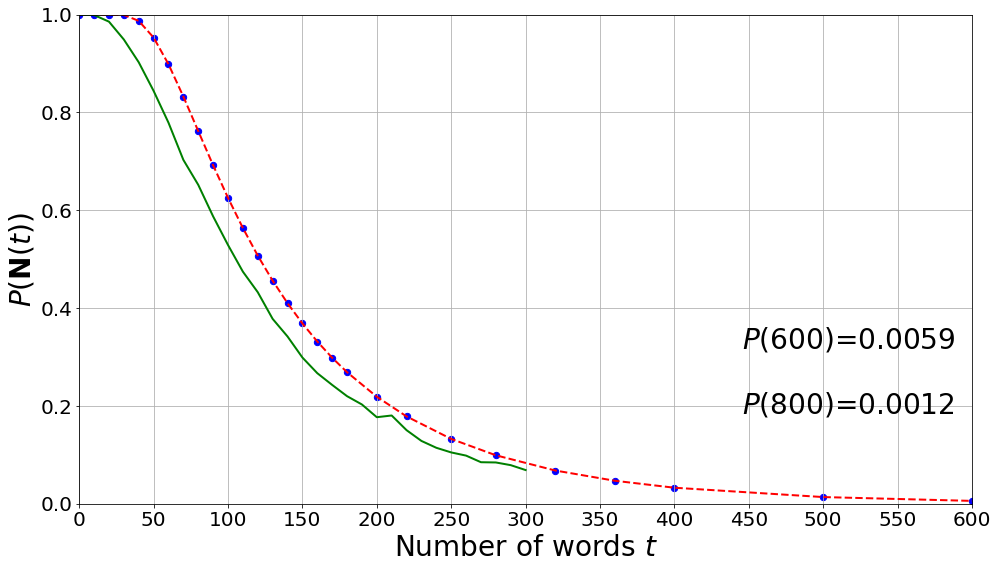}
		\caption{\textit{$\rho = 0.1$}.} \label{fig::rho_1}
	\end{subfigure}
	\hspace*{\fill} 
	\begin{subfigure}[t]{0.32\textwidth}
		\includegraphics[width=\linewidth]{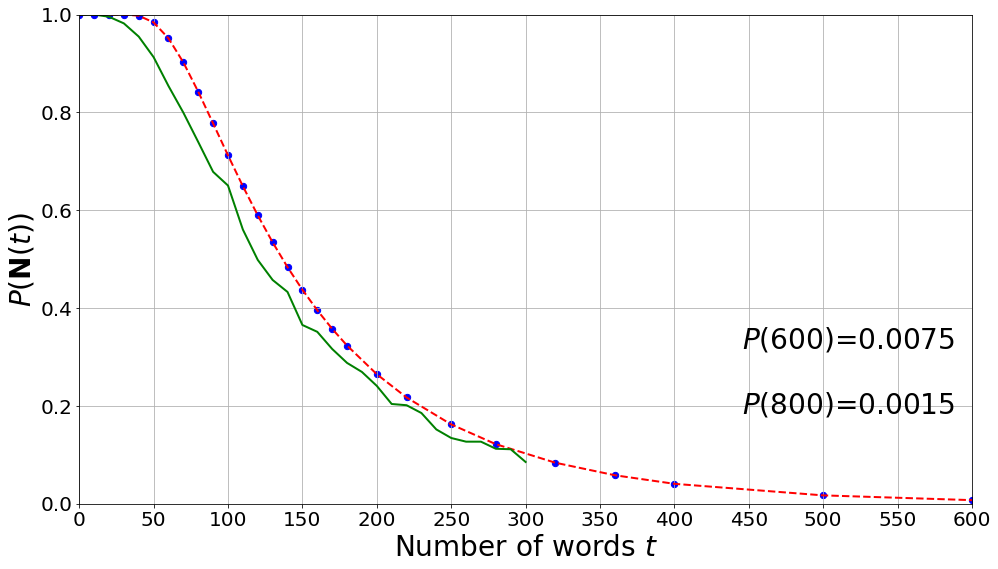}
		\caption{\textit{$\rho = 0.2$}.} \label{fig::rho_2}
	\end{subfigure}
	\bigskip
	\caption{Upper bound and simulation result of $P(t)$ with different transition probability $\rho$.}
	\label{fig::ptupperbound}
\end{figure*}

The red dashed line in Fig.~\ref{fig::ptupperbound} shows the upper bound on $P(t)$ when we set $m=2$, $P=\{0.4,0.3,0.2,0.1\}$ and $\rho = 0.05, 0.1$ and $0.2$. Simulation results are also shown in the figure. The simulation consists of repeating the following steps 2,000 times.
\begin{itemize}
	\item Generate a length-20,000 sequence with alphabet $\{11, 10, 01, 00\}$ (40,000 symbols from $\{0,1\}$) and corresponding probability distribution $P=\{0.4,0.3,0.2,0.1\}$. 
	\item Apply SLC direct shaping code to the source sequence. After the first $t$ words having been encoded, if a recurrence occurs during the encoding process, stop and declare a recurrence.
	\item Transmit the encoded sequence through a binary symmetric channel BSC($\rho$) with transition probability $\rho$.
	\item Decode the received noisy sequence while making sure the first $t$ symbols are decoded correctly (noise free).
	\item After the first $t$ symbols having been decoded, if a recurrence occurs during the decoding process, stop and declare a recurrence. 
\end{itemize}
The green line in Fig.~\ref{fig::ptupperbound} shows the fraction of experiments that stop because of recurrences.  We see that $P(t)$ decreases rapidly as $t$ increases. Other distributions and transition probabilities have also been simulated and they all produce qualitatively similar results. This indicates that if we make sure the first $t$ data words are decoded correctly for large enough $t$, for example by combining shaping coding with error correction coding, we can significantly reduce the likelihood of future error propagation. One possible approach is the bootstrap scheme in~\cite{BochererSteinSchulte}, which essentially uses a reverse concatenation architecture that combines systematic error correction codes and shaping codes, and has found application in optical data transmission.

\begin{remark}
	$(\mathbf{n}^e(t),\mathbf{n}^d(t))$ acts like a random walk in a $(2^{m-1}-1) \times (2^{m-1}-1)$ space. Thus Algorithm~\ref{algo::calculateq} can be used to numerically approximate a lower bound on $P_W$. \qed
\end{remark}

\subsection{Recurrence behavior of \textit{The Count of Monte Cristo}}
\begin{figure*}[t]
	\centering
	\begin{subfigure}[t]{0.32\textwidth}
		\includegraphics[width=\linewidth]{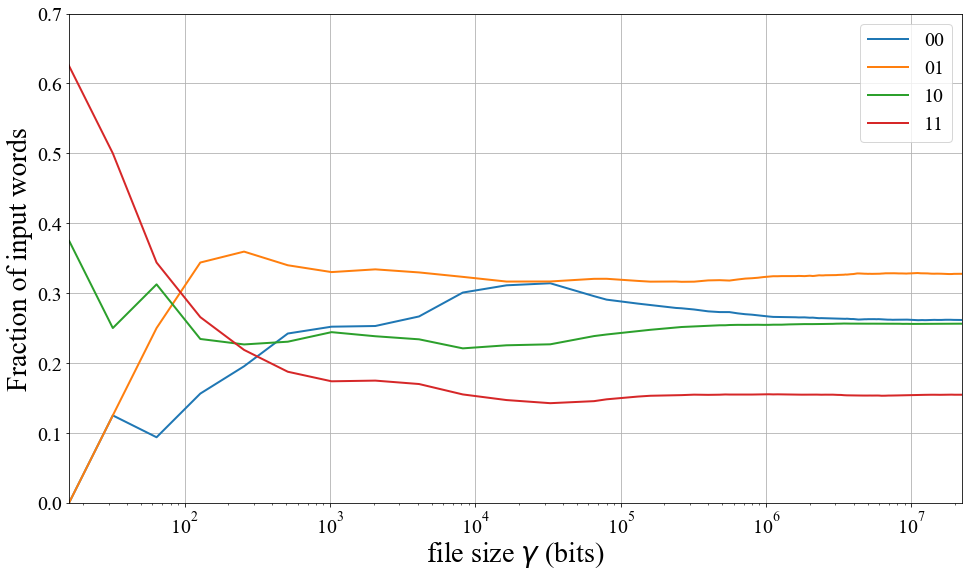}
		\caption{Length-2 dictionary.} \label{fig::fowlength2}
	\end{subfigure}
	\hspace*{\fill} 
	\begin{subfigure}[t]{0.32\textwidth}
		\includegraphics[width=\linewidth]{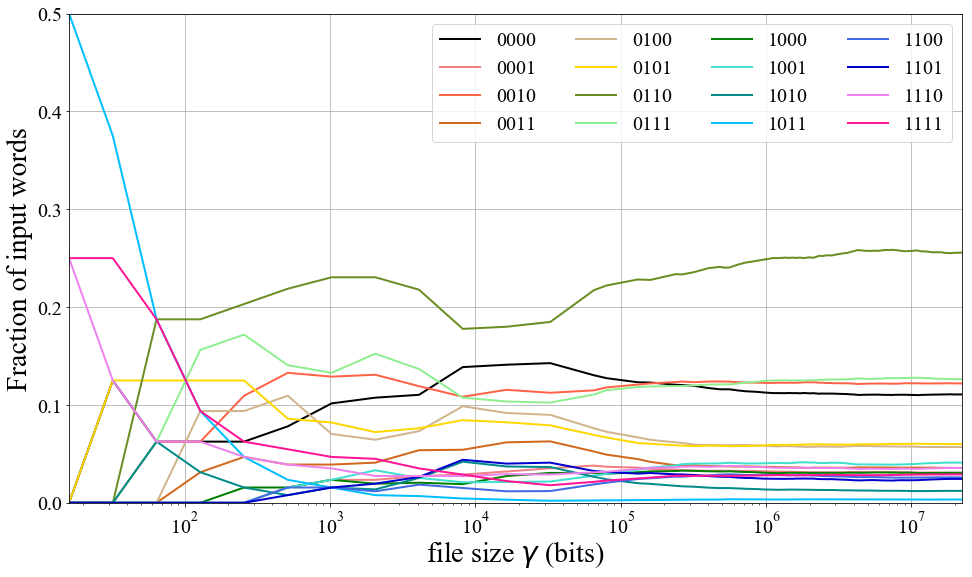}
		\caption{Length-4 dictionary.} \label{fig::fowlength4}
	\end{subfigure}
	\hspace*{\fill} 
	\begin{subfigure}[t]{0.32\textwidth}
		\includegraphics[width=\linewidth]{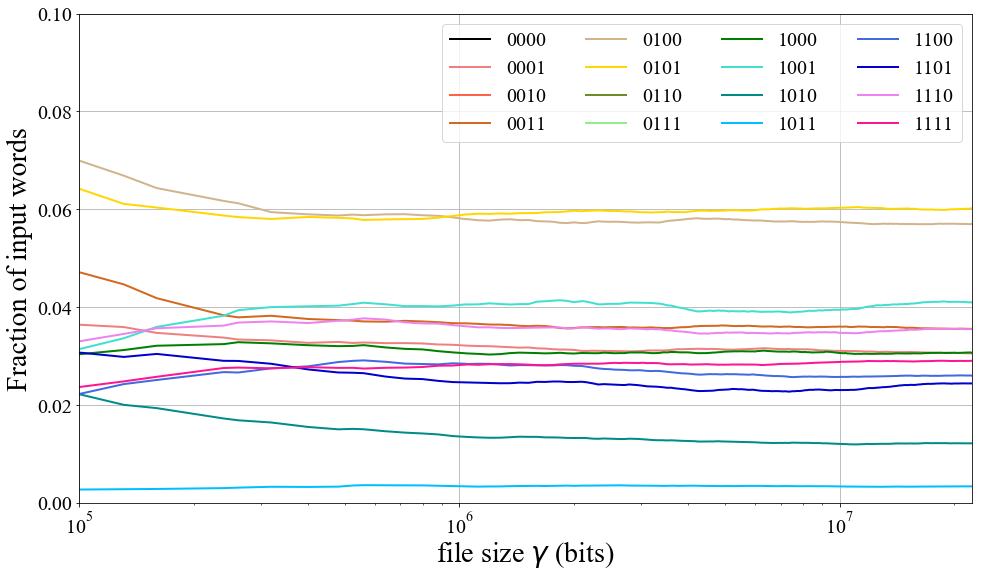}
		\caption{Length-4 dictionary detail.} \label{fig::fowlength4detail}
	\end{subfigure}
	\bigskip
	\caption{Fraction of words for segments of \textit{The Count of Monte Cristo}.}
	\label{fig::fractionofwords}
\end{figure*}

Here we study the sequence generated by applying length-2 and length-4 direct shaping codes to TCMC. The fractions of input words for segments of TCMC are shown in Fig.~\ref{fig::fractionofwords}. As shown in Fig.~\ref{fig::fowlength4} and~\ref{fig::fowlength4detail}, some words in the length-4 dictionary are not fully separated. This indicates that error propagation can happen throughout the entire encoding process. 

\begin{figure}[h]
	\centering
	\includegraphics[width=0.6\columnwidth]{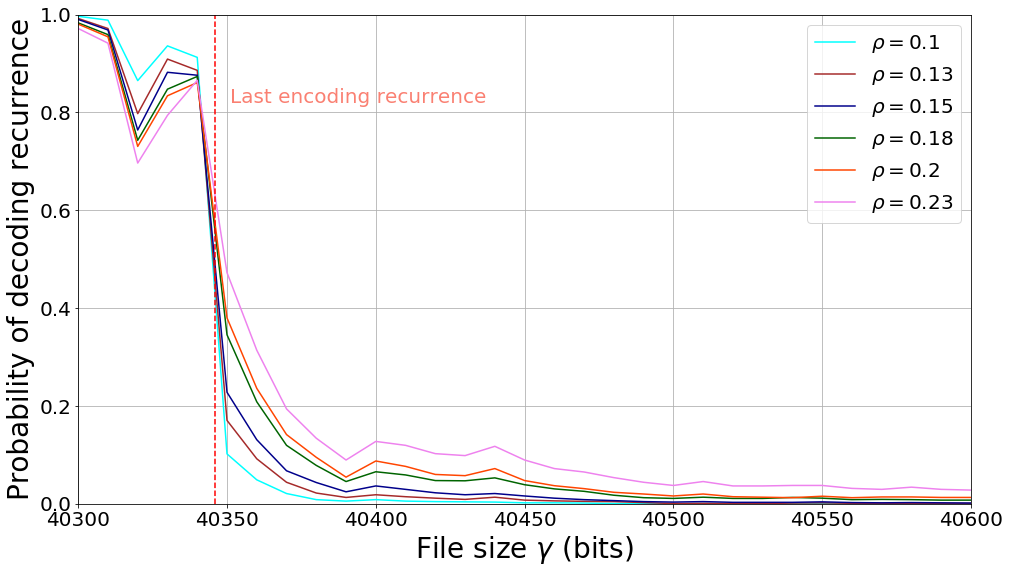}
	\caption{Probability of decoding recurrence when segments of \emph{The Count of Monte Cristo} are decoded correctly.}
	\label{fig::TCMC_decoding}
\end{figure}

We now examine the decoding recurrence probability of the length-2 direct shaping code. As shown in Fig.~\ref{fig::fowlength2}, the last encoding recurrence of a length-2 direct shaping code happens between words 00 and 11 at $\gamma = 40,346$ bits. We did the following simulation, which consists of repeating the following steps 2000 times.
\begin{itemize}
	\item Encode the first 10,000 bytes of \text{TCMC} (80,000 bits, or 40,000 input words).
	\item Transmit the encoded sequence through a binary symmetric channel BSC($\rho$)  with transition probability $\rho$.
	\item Decode the received noisy sequence while making sure the first $t$ symbols are decoded correctly (noise free).
	\item After the first $t$ symbols having been decoded, if a recurrence occurs during the decoding process, stop and declare a recurrence. 
\end{itemize}
The probability of decoding recurrence $P_d(t)$ with different $\rho$ is shown in Fig.~\ref{fig::TCMC_decoding}. As shown in the figure, $P_d(t)$ quickly converges to 0 when $t\geq40,346$ and the rate of convergence increases as $\rho$ decreases.

\section{Performance of Direct Shaping Codes}
\label{sec::performance}
	The performance of rate-constrained shaping codes was studied in~\cite{LiuJournal}. Given an i.i.d. source $\mathbf{X} = X_1,X_2\ldots$ with entropy $H$ and a variable-length mapping $\phi: \mathcal{X}^q \rightarrow \mathcal{Y}^*$ used as a shaping code, the relationship between the code expansion factor $f$, which is the ratio of the expected codeword length to the input word length, and the average cost was analyzed. In particular, the minimum achievable average cost associated with a fixed expansion factor was determined.
	
	In~\cite[Remark 1]{LiuJournal}, it was shown that direct shaping codes are shaping codes with parameters $q=1$ and $f=1$, where both the input and output processes have alphabet size $2^m$. The cost process $\mathbf{C} = \{C_t\}$ of a direct shaping code $\phi$ is
	\begin{equation}
	C_1 = c(\phi(X_1)), C_2=\frac{c(\phi(X_1,X_2))}{2},\ldots, C_t=\frac{c(\phi(X_1,X_2,\ldots X_t))}{t},
	\end{equation}
	and the asymptotic expected average cost is defined as 
	$C_{\infty}=\lim_{t\rightarrow \infty}E(C_t)$.
	In this section, we discuss the performance of direct shaping codes under the assumption of no errors. 
	The following theorem indicates the asymptotic expected average cost achievable by a direct shaping code.
	\begin{theorem}\label{performance:dsc}
		Given the distribution $P$ and cost vector $\mathcal{C}$, the asymptotic expected average cost $C_{\infty}$ of a direct shaping code is $\sum_{i}P_ic_i/m$.
	\end{theorem}
	
\begin{IEEEproof}
		We first assume that $P_1 > P_2>\ldots >P_{2^m}$. 
		Consider a sequence of i.i.d. random variables $X^{ij}_1,X^{ij}_2,\ldots$ with $P(X^{ij}_1=1)=P_i$, $P(X^{ij}_1=-1)=P_j$ and $P(X^{ij}_1=0)=1-P_i-P_j$. The random variable $X_k^{ij}$ corresponds to the change in distance between symbol $i$ and $j$ at time $k$. The expected value of $X^{ij}_1$ is $\mu^{ij}=P_i-P_j$. Define random variable $S^{ij}_t=\sum_{k=1}^{t}X^{ij}_k$ and note that $S^{ij}_t>0$ means $n_i(t)>n_j(t)$. By the strong law of large numbers, 
		\begin{equation}
		\label{equ::dictionarypairas}
		S^{ij}_t/t-\mu^{ij}\rightarrow 0 \quad \text{a.s.}.
		\end{equation} 
		For any $\epsilon>0$, $\{S^{ij}_t/t\}_0^\infty$ is within $\epsilon$ of $\mu^{ij}=P_i-P_j>0$ for all but finitely many $t$. This means for any two symbols $i>j$, $n_i(t)>n_j(t)$ for all but finitely many $t$. In other words for any $\epsilon,\delta >0$, there exists $T_1$ such that the probability of event $n_i(t) - n_j(t) > \epsilon$ for any $P_i>P_j$ and $t>T_1$ is greater than $1-\delta$. This indicates the dictionary is stable almost surely, i.e., after some time $T$, with probability greater than $1-\delta$, the dictionary is stable and a recurrence never occurs.
		
		Here we define another i.i.d. process $\{\widetilde{c}(X_t)\}$ with $\widetilde{c}(X) = c_i$ if $X= \mathbf{w_i}$. By the strong law of large numbers, we know that
		\begin{equation}
		\widetilde{C}_t= \frac{\widetilde{c}(X_1,X_2,\ldots, X_t)}{t} =  \frac{\sum_i\widetilde{c}(X_i)}{t}\rightarrow \sum P_ic_i \quad \text{a.s.}
		\end{equation}
		where we implicitly extend the definition of $\widetilde{c}$ to vectors additively.
		If after symbols $X_1^{t-1}$ the dictionary is stable, for the next input $X_t$ we have
		\begin{equation}
			\widetilde{c}(X_t) = c(\phi(X_1,X_2,\ldots X_t)) - c(\phi(X_1,X_2,\ldots X_{t-1})).
		\end{equation} 
		Since the dictionary is stable almost surely, for any $\delta, \epsilon >0$, there exists $T_1$ such that with probability greater than $1-\delta/2$, the dictionary is stable for any $t > T_1$ and there exists $T_2$ such that with probability greater than $1-\delta/2$, $|\widetilde{C}_t- \sum P_ic_i| < \frac{\epsilon}{2}$ for all $t > T_2$. Therefore with probability greater than $1-\delta$, for any $t > T' = \max\{T_1,T_2\}$, we have 
		\begin{equation}
		\label{equ::ctdivide}
		\begin{aligned}
		C_t &=\frac{c(\phi(X_1,X_2,\ldots X_t))}{t} \\
		&= \frac{c(\phi(X_1,X_2,\ldots X_{T'})) + c(\phi(X_1,X_2,\ldots X_{t})) - c(\phi(X_1,X_2,\ldots X_{T'}))}{t}\\
		&= \frac{c(\phi(X_1,X_2,\ldots X_{T'})) + \widetilde{c}(X_{T'+1}\ldots X_t)}{t}\\
		& = \frac{c(\phi(X_1,X_2,\ldots X_{T'})) -\widetilde{c}(X_1,X_2,\ldots X_{T'}) + \widetilde{c}(X_1,X_2,\ldots X_{t})}{t},
		\end{aligned}
		\end{equation}
		and
		\begin{equation}
		\label{equ::stabledictaverage}
			|\frac{\widetilde{c}(X_1,X_2,\ldots X_{t})}{t}- \sum P_i c_i| < \frac{\epsilon}{2}.
		\end{equation}
		Since the cost is bounded by $c_1$ and $c_{2^m}$, for the first $T'$ steps we have 
		\begin{equation}
		\label{equ::firstTprime}
		|c(\phi(X_1,X_2,\ldots X_{T'})) -\widetilde{c}(X_1,X_2,\ldots X_{T'}) |\leq T'(c_1 - c_{2^m}).
		\end{equation}
		Combining equations~(\ref{equ::ctdivide}),~(\ref{equ::stabledictaverage}) and~(\ref{equ::firstTprime}), we have
		\begin{equation}
		\begin{aligned}
		|C_t - \sum P_i c_i| &\leq |C_t - \widetilde{C}_t| + |\widetilde{C}_t - \sum P_i c_i|\\
		& < \frac{|c(X_1,X_2,\ldots X_{T'}) -\widetilde{c}(X_1,X_2,\ldots X_{T'})|}{t} +  \frac{\epsilon}{2}\\
		& \leq (c_1 - c_{2^m})\frac{T'}{t} +  \frac{\epsilon}{2}.
		\end{aligned}
		\end{equation}
		Let $T = \max \{T_1, T_2, \frac{2}{\epsilon}(c_1 - c_{2^m})T'\}$, then with probability greater than $1-\delta$, $|C_t -\sum P_i c_i| < \epsilon$ for any $t > T$. This indicates that
		\begin{equation} 
		C_t \rightarrow \sum P_ic_i \quad \text{a.s.}.
		\end{equation}
		Since $C_t$ is bounded by $c_1$ and $c_{2^m}$, by the dominated convergence theorem we have 
		\begin{equation}
		C_{\infty}=\lim_{t\rightarrow \infty}E(C_t) =\sum P_i c_i.
		\end{equation}
		When there exists some $i$ such that $P_i = P_{i+1}$, equation~(\ref{equ::dictionarypairas}) becomes
		\begin{equation}
			S^{ii+1}_t/t-0\rightarrow 0 \quad \text{a.s.}.
		\end{equation}
		This means that the almost sure stability condition is not satisfied. However, this doesn't affect the conclusion, because almost surely the input words $\mathbf{w_i}$ and $\mathbf{w_{i+1}}$ are mapped to $\mathbf{y_i}$ and $\mathbf{y_{i+1}}$ and $P(\mathbf{y_i}) = P(\mathbf{y_{i+1}}) = P_i$. This implies that the expected value of average cost satisfies
		$C_{\infty}=\lim_{t\rightarrow \infty}E(C_t) =\sum P_i c_i$.
		This completes  the proof.
\end{IEEEproof}

It was proved in~\cite{LiuJournal} that the minimum average cost of a rate-1 shaping code for costly channels with cost vector $\mathcal{C}$ and an i.i.d. source with entropy $H$ is $\sum_i \hat{P_i} c_i$, where
\begin{equation}
\hat{P_i} = \frac{2^{-\mu c_i}}{\sum_j 2^{-\mu c_j}}, \quad \mu\geq 0 \text{ and }-\sum_i \hat{P_i} \log_2 \hat{P_i} = H.
\end{equation}
This indicates that direct shaping codes are in general suboptimal. We have the following corollary.
\begin{cor}
	For an i.i.d. source with entropy $H$, SLC direct shaping codes are optimal with respect to average wear cost  if and only if
	\begin{equation}
	P_i =  \frac{2^{-\mu c_i}}{\sum_j 2^{-\mu c_j}}, \quad \mu\geq 0 \text{ and }-\sum_i P_i \log_2 P_i = H.
	\end{equation}
\end{cor}

\section{Conclusion}
In this paper, we studied shaping codes to reduce programming wear when writing structured data to flash memory. We first reviewed so-called direct shaping codes for SLC flash memory. Using a page-oriented, programming cost model, we  then extended this technique to MLC flash memory. The performance of these shaping codes was empirically evaluated on English and Chinese language text. Then we examined the error propagation behavior of direct shaping codes. We showed that by making sure the first $t$ data words are decoded correctly for large enough $t$, we can significantly reduce the likelihood of future error propagation.
Finally, we derived the asymptotic average cost of a direct shaping code.  The results indicate that direct shaping codes are in general suboptimal  and can only achieve the minimum average cost for rate-1 codes if the source distribution satisfies a specific condition.  

\section*{Acknowledgment}
The authors thank Professor Jason Schweinsberg for discussions on multi-dimensional random walks in cones. This work was supported in part by National Science Foundation (NSF) Grant CCF-1619053, the Center for Memory and Recording Research at UC San Diego, and Toshiba Corporation.

	\begin{appendices}
		\section{PROOF OF LEMMA~\ref{lemma::generalupperbound}}
		\label{appen::generalupperbound}
		\begin{IEEEproof}
			From the union bound, we have
			\begin{equation}
			\small
			\label{func::first_union_bound}
			\begin{split}
			P_W&=\text{Pr}\bigg{\{}\left(\bigcup_{i=1}^{2^{m-1}-1}\{i,i+1\}^e\right)\cup \left(\bigcup_{i=1}^{2^{m-1}-1}\{i,i+1\}^d\right)\bigg{\}}\leq \text{Pr}\bigg{\{}\bigcup_{i=1}^{2^{m-1}-1}\{i,i+1\}^e\bigg{\}} + \text{Pr}\bigg{\{}\bigcup_{i=1}^{2^{m-1}-1}\{i,i+1\}^d\bigg{\}}.
			\end{split}
			\end{equation}
			Using the fact that  $\{i,i+1\}^e$ and $\{i+2,i+3\}^e$ are independent, we have
				\begin{equation}\label{func::peunionbound}
				\begin{split}
				\text{Pr}\bigg{\{}\bigcup_{i=1}^{2^m-1}\{i,i+1\}^e\bigg{\}}&=\text{Pr}\bigg{\{}\left(\bigcup_{i=1}^{2^{m-1}}\{2i-1,2i\}^e\right)\bigcup\left(\bigcup_{i=1}^{2^{m-1}-1}\{2i,2i+1\}^e\right)\bigg{\}}\\
				&\leq\text{Pr}\bigg{\{}\bigcup_{i=1}^{2^{m-1}}\{2i-1,2i\}^e\bigg{\}}+\text{Pr}\bigg{\{}\bigcup_{i=1}^{2^{m-1}-1}\{2i,2i+1\}^e\bigg{\}}\\
				&=2-\text{Pr}\bigg{\{}(\bigcup_{i=1}^{2^{m-1}}\{2i-1,2i\}^e)^C\bigg{\}}-\text{Pr}\bigg{\{}(\bigcup_{i=1}^{2^{m-1}-1}\{2i,2i+1\}^e)^C\bigg{\}}\\
				&=2-\text{Pr}\bigg{\{}(\bigcap_{i=1}^{2^{m-1}}\{\oline{2i-1,2i}\}^e)\bigg{\}}-\text{Pr}\bigg{\{}(\bigcap_{i=1}^{2^{m-1}-1}\{\oline{2i,2i+1}\}^e)\bigg{\}}\\
				&=2-\prod_{i=1}^{2^{m-1}}\left(1-\frac{P_{2i}}{P_{2i-1}}^{N_{2i-1}(t)}\right)-\prod_{i=1}^{2^{m-1}-1}\left(1-\frac{P_{2i+1}}{P_{2i}}^{N_{2i}(t)}\right).
				\end{split}
				\end{equation}
			Similarly, for the decoding process we have
				\begin{equation}
				\label{func::pdunionbound}
				\begin{aligned}
				&\text{Pr}\bigg{\{}\bigcup_{i=1}^{2^m-1}\{i,i+1\}^d\bigg{\}}\leq 2-\prod_{i=1}^{2^{m-1}}\left(1-\frac{P^d_{2i}}{P^d_{2i-1}}^{N_{2i-1}(t)}\right)-\prod_{i=1}^{2^{m-1}-1}\left(1-\frac{P^d_{2i+1}}{P^d_{2i}}^{N_{2i}(t)}\right).
				\end{aligned}
				\end{equation}
			Combining equations~(\ref{func::first_union_bound}),~(\ref{func::peunionbound}), and~(\ref{func::pdunionbound}) completes the proof.
		\end{IEEEproof}
\end{appendices}

\begin{thebibliography}{99}
	\bibitem{BochererSteinSchulte}G. B\"{o}cherer, F. Steiner, and P. Schulte, ``Bandwidth efficient and rate-matched low-density parity-check coded modulation,'' \emph{IEEE Trans. Commun.}, vol. 63, no. 12, pp. 4651--4665, Dec. 2015.
	\bibitem{Chee} Y. M. Chee, H. M. Kiah, A. J. H. Vince, V. K. Vu, and E. Yaakobi, ``Coding for write l-step-up memories'', in \emph{Proc. IEEE Int. Symp. Inf. Theory (ISIT)}, Paris, France, 2019, pp. 1597-1601.
	\bibitem{Cover}T. Cover, ``Enumerative source encoding,'' \emph{IEEE Trans. Inf. Theory}, vol. 19, no. 1, pp. 73-77, Jan. 1973.
	\bibitem{Durrett} R. Durrett, \emph{Probability: Theory and Examples}, 5th ed. Cambridge university press, 2019.
	\bibitem{Golin}M. J. Golin and G. Rote, ``A dynamic programming algorithm for constructing optimal prefix-free codes with unequal letter costs," \emph{IEEE Trans. Inf. Theory}, vol. 44, no. 5, pp. 1770-1781, Sep. 1998.
	\bibitem{Guazzo}M. Guazzo, ``A general minimum-redundancy source-coding algorithm," \emph{IEEE Trans. Inf. Theory}, vol. 26, no. 1, pp. 15-25, Jan. 1980.
	\bibitem{Jagmohan} A. Jagmohan, M. Franceschini, L. A. Lastras-Monta\~{n}o and J. Karidis, ``Adaptive endurance coding for NAND Flash," in \emph{Proc. IEEE GLOBECOM Workshops}, Dec. 2010, pp. 1841--1845.
	\bibitem{Karp}  R. Karp, ``Minimum-redundancy coding for the discrete noiseless channel," \emph{IRE IEEE Trans. Inf. Theory}, vol. 7, no. 1, pp. 27-38, Jan. 1961.
	\bibitem{Krause} R. M. Krause, ``Channels which transmit letters of unequal duration," \emph{Inf. Contr.}, vol. 55, pp. 13-24, 1962. 
	\bibitem{Li}J. Li, K. Zhao, X. Zhang, J. Ma, M. Zhao, and T. Zhang, ``How much can data compressibility help to improve NAND flash memory lifetime?,'' in \emph{Proc. 13th USENIX Conf.  File and Storage Technologies} (FAST'15), USENIX Assoc., Berkeley, CA, Feb. 16-19, 2015, pp. 227-240.
	\bibitem{LiuJournal} Y. Liu, P. Huang, A. W. Bergman, and P. H. Siegel,``Rate-constrained shaping codes for structured source,'' \emph{arXiv preprint arXiv:2001.02748}, 2020.
	\bibitem{LiuML} Y. Liu, S. Wu, and P. H. Siegel, ``Bad Page Detector for NAND Flash Memory'', \emph{11th Annual Non-Volatile Memories Workshop (NVMW)}, La Jolla, CA, Mar. 8-10, 2020.  
	\bibitem{Nussbaum} R. D. Nussbaum, ``Convexity and Log Convexity for the Spectral Radius,'' \emph{Linear Algebra and its Applications}, 73:59-122, 1986.
	\bibitem{Rivest}  R. L. Rivest and A. Shamir, ``How to reuse a write-once memory,'' \textit{Inform. and Contr.}, vol. 55, no. 1-3, pp. 1-19, Dec. 1982.
	\bibitem{Shannon} C. E. Shannon, ``A mathematical theory of communication, Part I, Part II," \emph{Bell Syst. Tech. J,} vol. 27, pp. 379--423, 1948. 
	\bibitem{Sharon}E. Sharon, et al., ``Data Shaping for Improving Endurance and Reliability in Sub-20nm NAND,'' presented at \emph{Flash Memory Summit}, Santa Clara, CA, Aug. 4-7, 2014.
\end{thebibliography}
\end{document}